\documentclass[twocolumn,english,preprintnumbers,amsmath,amssymb,pre]{revtex4}

\usepackage{graphicx}
\usepackage{color}

\begin{document}

\title{Nuclear quantum effects in graphane}
\author{Carlos P. Herrero}
\author{Rafael Ram\'irez}
\affiliation{Instituto de Ciencia de Materiales de Madrid,
         Consejo Superior de Investigaciones Cient\'ificas (CSIC),
         Campus de Cantoblanco, 28049 Madrid, Spain }
\date{\today}

\begin{abstract}
Graphane is a quasi-two-dimensional material consisting of
a single layer of fully hydrogenated graphene, with a C:H ratio
of 1. We study nuclear quantum effects in the so-called
chair-graphane by using path-integral molecular dynamics (PIMD)
simulations.   The interatomic interactions are modeled by a
tight-binding potential model fitted to density-functional
calculations. Finite-temperature properties are studied in
the range from 50 to 1500~K.  To assess the magnitude of nuclear
quantum effects in the properties of graphane, classical molecular
dynamics simulations have been also performed. These quantum
effects are significant in structural properties such as interatomic
distances and layer area at finite temperatures. The in-plane
compressibility of graphane is found to be about twice larger than
that of graphene, and at low temperature it is 9\% higher than
the classical calculation. The thermal expansion coefficient
resulting from PIMD simulations vanishes in the zero-temperature
limit, in agreement with the third law of Thermodynamics.   \\
\end{abstract}

\maketitle

\section{Introduction}

Carbon-based materials have been intensively investigated in
recent years, in particular those consisting of two-dimensional (2D) 
layers \cite{ge07,wo14,me16,ca09b,ca18,ya19}.
In this context, hydrogenated graphene (called graphane)
is a quasi-2D structure of C atoms ordered in a buckled honeycomb 
lattice covalently bonded to H atoms.
The most studied conformer of graphane is the so-called
chair-graphane, where H atoms alternate in a chairlike arrangement
on both sides of the carbon layer \cite{so07,we11}. This graphane
configuration is studied in this paper.
There exist also boat- and washboard-graphane \cite{ca10},
which will not be considered here.

Graphane can be reversibly obtained by hydrogen chemisorption 
on pure graphene \cite{el09}, which causes a rearrangement
of the chemical bonds and angles in the honeycomb lattice 
of graphene. Each C atom is bound to an H neighbor,
changing its orbital hybridization from sp$^2$ to sp$^3$, 
and the planar configuration of graphene
is modified into an out-of-plane buckled structure.
Graphane is a wide band-gap semiconductor, where appreciable 
spin polarization can be achieved by the creation of 
domains of H vacancies and CH divacancies \cite{sa10b}.
Moreover, the presence of impurities such as Li atoms or metal 
dopants may significantly affect its electronic and magnetic 
properties \cite{wa16b,ma17b,en13}.

A deep comprehension of structural and thermal properties 
of 2D systems is a challenging problem in modern statistical 
physics \cite{sa94,ne04,ta13},
which has been mainly discussed in the field of biological
membranes and soft condensed matter \cite{fo08,ta13}.
However, the large complexity of these systems makes it
difficult to devise microscopic models on the basis of
realistic interatomic interactions.
2D carbon-based materials provide us with model systems where
atomic-scale studies are possible, allowing for a deeper
understanding of the physical properties of this type
of systems \cite{po12b,fo13,wa16,he18}.

At finite temperatures, thermally excited ripples appear and
distort the lattice of 2D materials.   
It has been suggested that in graphane
the aspect of these ripples may be different from those in
graphene. This could be a consequence of the fact that in graphane
the thermal energy can be accommodated on the in-plane bending modes
(involving C-C-C bond angles in the buckled configuration),
instead of leading to significant out-of-plane fluctuations, as
happens in graphene \cite{co12,ga14}.

In several electronic-structure calculations of graphane presented
in the literature, even though they are based on precise {\em ab-initio} 
quantum mechanical methods, atomic nuclei are described as classical 
particles \cite{so07,ca10,le10,zh13,ch14},
so that some quantum effects such as zero-point motion are
not included in the calculation.
Finite-temperature properties of graphane have been also studied
by molecular dynamics simulations using {\em ab-initio} \cite{ch14}
and empirical interatomic potentials \cite{co12,li13}.
In these simulations, atomic nuclei were also treated as 
classical particles.

Nuclear quantum effects may be important for vibrational and electronic
properties of relatively light elements like carbon, and even more for
hydrogen, especially at low temperatures.
To take into account the quantum nature of the nuclei,
path-integral (Monte Carlo and molecular dynamics) simulations
are especially adequate, since the nuclear degrees of freedom can be
efficiently quantized, allowing one to study quantum and thermal
fluctuations at finite temperatures \cite{gi88,ce95}.
This procedure permits to perform quantitative studies
of anharmonic effects in condensed matter \cite{he95,ra11}.

In this paper we use the path-integral molecular dynamics (PIMD) 
method to study the influence of nuclear quantum dynamics on
structural, vibrational, and thermal properties of graphane at 
temperatures from 50 to 1500~K. 
The interatomic interactions are described by an efficient 
tight-binding (TB) Hamiltonian, developed on the basis of 
density-functional calculations.
We consider simulation cells of different sizes, 
as finite-size effects can be relevant for some variables, 
such as the atomic delocalization in the out-of-plane 
direction \cite{ga14,lo16}.
Path-integral methods similar to that employed here
have been applied before to study nuclear quantum effects 
in carbon-based materials as diamond \cite{he00c,ra06,he07}, and
more recently in graphene \cite{br15,ha18,he16}.
The adsorption and diffusion of H on graphene has been also studied 
by using this kind of techniques \cite{da14,he09a}.
Moreover, nuclear quantum effects have been analyzed
earlier by using a combination of density-functional theory and
a quasi-harmonic approximation for vibrational modes
in graphane \cite{hu13b}.

 The paper is organized as follows. In Sec.\,II we describe the
computational techniques employed here: PIMD method, tight-binding 
procedure, and calculation of atomic mean-square displacements.
In Sec.\,III we present results for the internal energy of graphane,
with emphasis on its constituent parts, i.e., kinetic and potential 
energy.  Results for structural properties are given in Sec.~\,IV
(interatomic distances) and Sec.~\,V (orientation of the C--H bonds).
In Sec.\,VI we study the atomic motion, as visualized from
mean-square displacements in the in-plane and out-of-plane directions.
Data for the layer area  and the in-plane compressibility of 
graphane are given in Secs.~\,VII and VIII, respectively.
The paper closes in Sec.\,IX with a summary of the main results.

\section{Computational Method}

\subsection{Path-integral molecular dynamics}

We use the PIMD method to obtain equilibrium properties of graphane
at several temperatures.
This procedure, based on the Feynman path-integral formulation of 
statistical mechanics \cite{fe72}, is a nonperturbative approach
suitable to study finite-temperature properties of many-body 
quantum systems.
It profits from the fact that the partition function of a quantum 
system may be expressed in a way formally equivalent to that of a 
classical one, obtained by replacing each quantum particle 
by a ring polymer formed by $N_{\rm Tr}$ (Trotter number) classical 
particles, linked by harmonic springs \cite{fe72,kl90,ce95}.
Details on this simulation technique can be found
elsewhere \cite{ch81,gi88,tu10,he14}.

We employ the molecular dynamics method to sample the configuration
space of the classical isomorph of our quantum system.  
The dynamics in this computational procedure is artificial, 
since it does not represent
the actual quantum dynamics of the real particles.
Nevertheless, it is well-suited for effectively sampling the
many-body configuration space, yielding accurate values for 
time-independent equilibrium properties of the quantum system
under consideration.
The calculations presented here were performed in the 
isothermal-isobaric ensemble, where we fix the number of atoms 
($N$ pairs C--H), the in-plane applied stress (here $\tau = 0$), 
and the temperature ($T$).
The stress $\tau$ in the $(x, y)$ plane, with units of 
force per unit length, coincides with the so-called 
mechanical or frame tension \cite{ra17,fo08,sh16,he18b}.
We have used effective algorithms for carrying out the PIMD 
simulations in the $N \tau T$ ensemble, as those presented in the
literature \cite{ma99,tu10}. Specifically, staging variables 
have been employed to define the bead coordinates, and
the constant-temperature ensemble was achieved by coupling chains
of four Nos\'e-Hoover thermostats to each staging variable.
Moreover, a chain of four barostats was coupled to the in-plane 
area of the simulation cell to give the required pressure, 
$\tau = 0$ \cite{tu10,he14}.

\begin{figure}
\vspace{-0.6cm}
\includegraphics[width=7cm]{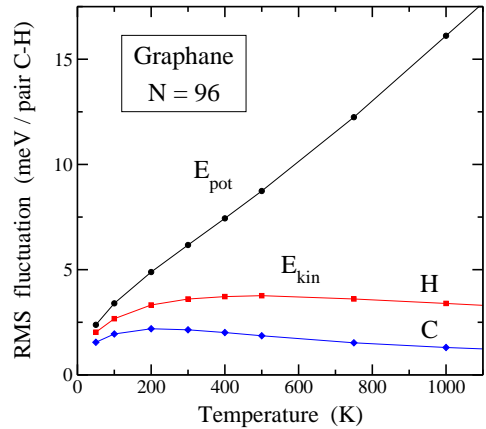}
\vspace{-0.5cm}
\caption{RMS fluctuations of the kinetic and potential energy
for graphane vs the temperature, as derived from PIMD simulations.
Circles: potential energy; squares: kinetic energy of hydrogen;
diamonds: kinetic energy of carbon.
Error bars are smaller than the symbol size.
Lines are guides to the eye.
}
\label{f1}
\end{figure}

The equations of motion have been integrated by using the reversible 
reference system propagator algorithm (RESPA), which permits us to 
define different time steps for the integration of the slow and fast
degrees of freedom \cite{ma96}.
The time step employed for the dynamics related to interatomic forces
was $\Delta t$ = 0.5 fs, which turned out to be suitable for the 
atomic masses and temperatures considered here, and provided good
convergence for the studied variables.
For the evolution of the fast dynamical variables, including
thermostats and harmonic bead interactions, we employed a
time step $\delta t = \Delta t/4$, as in previous 
simulations \cite{he06}.
The kinetic energy, $E_{\rm kin}$, has been calculated by means of 
the so-called virial estimator, which has a statistical uncertainty
smaller than the potential energy of the system, especially
at high temperatures \cite{he82,tu10}.
This can be seen in Fig.~1, where we have plotted the root mean-square
(RMS) fluctuations of the kinetic ($E_{\rm kin}$) and potential 
($E_{\rm pot}$) energy as a function of temperature. 
The RMS fluctuations of $E_{\rm pot}$
increase for rising $T$, while those of $E_{\rm kin}$ reach 
a maximum and then decrease at high $T$ for both H and C.

The configuration space of graphane for simulation cells
including $2N$ atoms ($N$ pairs C--H, with $N$ from 24 to 216), 
has been sampled at temperatures between 50 and 1500~K.
For the smallest size, $N$ = 24, PIMD simulations at $T$ = 25 K were 
performed. To compare with the results of our quantum simulations, 
some classical molecular dynamics (MD) simulations of graphane 
have been also carried out.  In our context, this is achieved 
by setting the Trotter number $N_{\rm Tr}$ = 1.
In the quantum simulations, $N_{\rm Tr}$ was taken proportional to 
the inverse temperature, so that $N_{\rm Tr} \, T$ = 6000~K. 
This choice keeps roughly constant the precision associated 
to the finite values of $N_{\rm Tr}$ at different 
temperatures \cite{he06}.

\begin{figure}
\vspace{-2.5cm}
\includegraphics[width=7cm]{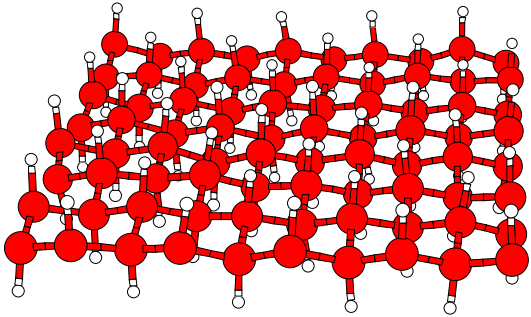}
\vspace{-1.5cm}
\caption{Snapshot taken from of a simulation of graphane
(96 C + 96 H) at $T = 300$~K.
Large dark and small white circles represent carbon and hydrogen
atoms, respectively.
}
\label{f2}
\end{figure}

We considered rectangular simulation cells with similar side length
in the $x$ and $y$ directions of the $(x, y)$ reference plane
($L_x \approx L_y$), for which
periodic boundary conditions were assumed.
In the out-of-plane $z$-direction, we have free boundary conditions,
so that C and H atoms can unrestrictedly move, simulating 
a free-standing graphane layer.
For a given temperature, a typical simulation run consisted of
$10^5$ PIMD steps for system equilibration, followed by
$2 \times 10^6$ steps for the calculation of ensemble average properties,
except for the cell size $N = 216$, for which the trajectories
included $8 \times 10^5$ steps.
In Fig.~2 we present a view of a graphane configuration
obtained in our simulations at $T = 300$~K.
In this picture, red and white circles represent C and H atoms,
respectively.

For comparison with our results for graphane, we have performed 
some PIMD simulations of graphene with the same TB potential as that
used for graphane (see below).  
For graphene, we employed cells consisting of $N$ carbon atoms, 
with $N$ = 96 and 216.
Moreover, some simulations were carried out for a single
H impurity on graphene, similar to those presented 
earlier \cite{he09a}. In the present case, however, these 
simulations were performed in the isothermal-isobaric ensemble, as
those of graphane, vs the constant in-plane area simulations in
Ref.~\cite{he09a}.

\subsection{Tight-binding procedure}

The calculations presented here have been carried out within the adiabatic
(Born-Oppenheimer) approximation, which permits to define a 
potential-energy surface for the nuclear coordinates.
A relevant point in the PIMD procedure is a satisfactory description 
of the interatomic interactions, which should be as realistic as possible.
Using density functional or Hartree-Fock based self-consistent 
potentials requires computational resources that would considerably restrict
the size of the manageable simulation cells and/or the number of
accessible PIMD steps.
Thus, we obtain the Born-Oppenheimer surface for the
nuclear dynamics from an efficient tight-binding Hamiltonian,
based on density functional calculations \cite{po95}.

The capability of TB methods to accurately describe various properties
of molecules and solids was reviewed by Goringe {\em et al.} \cite{go97}.
We have checked the predictions of this TB potential for well-known
frequencies of C--H vibrations in small molecules.
For example, for CH$_4$
it predicts in a harmonic approximation frequencies of 3100 and
3242 cm$^{-1}$ for C--H modes with symmetry $A_1$ and $T_2$,
respectively \cite{he06}, to be compared
with values of 2917 and 3019 cm$^{-1}$ obtained from vibrational
spectroscopy \cite{jo93}.
If one considers the anharmonic shift associated to these modes
(usually towards lower frequencies), the accord is acceptable.
A detailed study of vibrational frequencies in hydrocarbon
molecules obtained with this TB potential, taking into account
mode anharmonicities, was presented elsewhere \cite{lo03,bo01}.
We have used earlier this TB Hamiltonian to describe
carbon-hydrogen interactions in diamond \cite{he06,he07},
graphite \cite{he10b}, and graphene \cite{he09a}.

\begin{figure}
\vspace{-0.6cm}
\includegraphics[width=7cm]{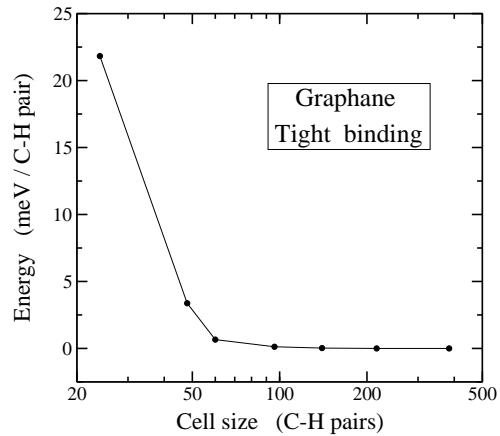}
\vspace{-0.5cm}
\caption{Convergence of the potential energy of graphane for
different cell sizes ($N$). The zero of energy is taken for
$N = 384$. Note the logarithmic scale in the horizontal axis.
}
\label{f3}
\end{figure}

An advantage of the combination of path integrals with electronic structure
methods is that both electrons and atomic nuclei are treated quantum
mechanically, so that phonon-phonon and electron-phonon interactions
are directly taken into account in the simulation.
For the reciprocal-space sampling we have employed only the $\Gamma$
point (${\bf k} = 0$), as the main effect of using a larger ${\bf k}$
set is a nearly constant shift in the total energy, with little influence
in the calculation of energy differences.
In Fig.~3 we present the convergence of the potential energy of 
graphane for different cell sizes $N$. The data points
correspond to $E_{\rm pot}$ obtained with the TB model for the 
minimum-energy configuration (classical, $T = 0$).

\subsection{Mean-square displacements}

An interesting application of PIMD simulations is the study of
atomic delocalization in three-dimensional (3D) space 
at finite temperatures. 
This includes a thermal (classical) delocalization,
and another associated to the quantum nature of atomic nuclei, 
which may be assessed by the extension of the quantum paths
sampled in the simulations.
For a path of nucleus $i$ ($i = 1, ..., 2N$),
we define the {\em centroid} (center of mass) as
\begin{equation}
   \overline{\bf r}_i = \frac{1}{N_{\rm Tr}} 
          \sum_{j=1}^{N_{\rm Tr}} {\bf r}_{ij} \; ,
\label{centr}
\end{equation}
where ${\bf r}_{ij} \equiv (x_{ij}, y_{ij}, z_{ij})$ is the
position of bead $j$ in the associated ring polymer.

The mean-square displacement $(\Delta x)^2_i$ of nucleus
$i$ in the $x$ direction along a PIMD simulation 
run is defined as
\begin{equation}
 (\Delta x)^2_i =  \frac{1}{N_{\rm Tr}}  \sum_{j=1}^{N_{\rm Tr}}
    \left<  ( x_{ij} - \left< \overline{x}_i \right>)^2 \right>  \, .
\label{deltax2b}
\end{equation}
In this expression $\overline{x}_i$ is the instantaneous $x$-coordinate 
of the centroid of atom $i$, and $\left< \overline{x}_i \right>$ is the
average value along a simulation run, which corresponds to the
observable $x$-coordinate of the atomic position.  Thus,
$\left<  ( x_{ij} - \left< \overline{x}_i \right>)^2 \right>$ is
the mean-square displacement (MSD) of the coordinate $x_{ij}$ 
of bead $j$ with respect to the average centroid 
$\left< \overline{x}_i \right>$ ($j$ = 1, ..., $N_{\rm Tr}$).
Hence, $(\Delta x)^2_i$ is an average of these displacements for the
beads associated to nucleus $i$, corresponding to the observable MSD
of the atomic coordinate.

The spread of the paths associated to an atomic nucleus
can be measured by the mean-square {\em radius-of-gyration} of
the ring polymers, with an $x$ component \cite{gi88,gi90}:
\begin{equation}
  Q_{x,i}^2 = \frac{1}{N_{\rm Tr}}   \sum_{j=1}^{N_{\rm Tr}}
           \left<  (x_{ij} - \overline{x}_i)^2 \right>    \, .
\label{qxi2}
\end{equation}
Notice the difference between the r.h.s. of Eqs.~(\ref{deltax2b})
and (\ref{qxi2}): in the former one has an average of the centroid
position over the whole trajectory, i.e. $\left< \overline{x}_i \right>$,
while in the latter there appears the instantaneous value 
$\overline{x}_i$ for each configuration.

At finite temperatures ($T > 0$~K), the observable spatial delocalization 
$(\Delta x)^2_i$ of nucleus $i$ in the $x$ direction contains, 
along with $Q_{x,i}^2$, another term which takes into account 
classical-like motion of the centroid coordinate $\overline{x}_i$:
\begin{equation}
    (\Delta x)^2_i = C_{x,i}^2 + Q_{x,i}^2  \, ,
\label{deltax2c}
\end{equation}
where
\begin{equation}
   C_{x,i}^2 =  \langle  \overline{x}_i^2 \rangle -
                \langle  \overline{x}_i \rangle^2  \, .
\label{cxi2}
\end{equation}
$C_{x,i}^2$ is the MSD of the centroid of nucleus $i$, and the 
quantum component $Q_{x,i}^2$ is the average MSD of the path (beads 
in the ring polymer) with respect to the instantaneous centroid. 
$C_{x,i}^2$ is a semiclassical thermal contribution
to $(\Delta x)^2_i$, as at high temperature it converges to
the MSD of a classical model, where the
quantum paths converge to single points ($Q_{x,i}^2 \to 0$).
In the limit $T \to 0$, $C_{x,i}^2$ vanishes and
$Q_{x,i}^2$ corresponds to zero-point motion of nucleus $i$.
For each atomic species (H and C),
we will present below results for $(\Delta x)^2_{\rm H}$  and
$(\Delta x)^2_{\rm C}$ calculated 
as averages for $N$ atoms in the simulation cell.
For example, for hydrogen we have
\begin{equation}
 (\Delta x)^2_{\rm H} = \frac{1}{N} \sum_{i=1}^{N} (\Delta x)^2_i  \, ,
\end{equation}
and similarly for $Q^2_{x,{\rm H}}$ and $C^2_{x,{\rm H}}$.
For the $y$ and $z$ directions we have similar expressions to those
given above for the $x$ direction.

\section{Energy}

In this section we present and discuss the internal energy 
of graphane, obtained in our isothermal-isobaric ensemble 
for $\tau = 0$ and several temperatures.
At $T = 0$ we find in a classical approach 
a graphane layer composed of two planar sheets of C atoms
(sublattices A and B) separated by 0.464 \AA, and two sheets
of H atoms on both sides at a distance of 1.126 \AA\ from the
nearest C atoms. Thus, the distance between H planes is 2.716 \AA.
This corresponds to a graphane layer with fixed atoms on their
equilibrium sites without spatial delocalization, giving the minimum
energy $E_0 = -57.0393$~eV/(C--H pair), taken as a reference for 
our calculations at finite temperatures.

In a quantum description of the atomic nuclei, the low-temperature 
limit displays in-plane and out-of-plane atomic fluctuations due to 
zero-point motion, and the C and H sheets are not strictly planar.
Moreover, anharmonicity of out-of-plane vibrations in the $z$-direction
gives rise to a small zero-point expansion, giving a distance between
H planes of 2.739 \AA, i.e. a dilation of 0.023 \AA.

\begin{figure}
\vspace{-0.6cm}
\includegraphics[width=7cm]{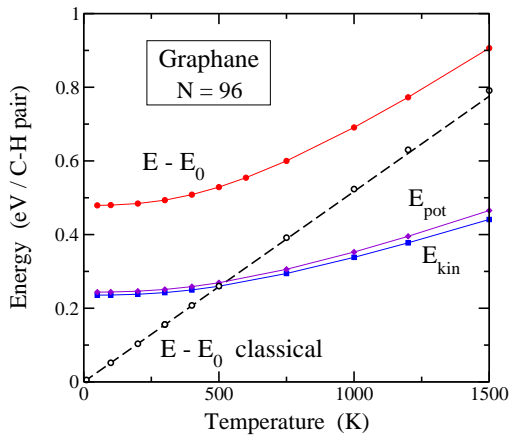}
\vspace{-0.5cm}
\caption{
Internal energy per C-H pair vs temperature, as derived from classical
(open circles) and PIMD simulations (solid circles) of graphane.
Solid squares and diamonds represent the kinetic and potential
energy obtained in the quantum simulations, respectively.
The dashed line corresponds to the classical limit of the vibrational
energy per C--H pair in a harmonic approximation:
$E_{\rm vib}^{\rm cl} = 6 k_B T$.
Solid lines are guides to the eye.
Error bars are less than the symbol size.
}
\label{f4}
\end{figure}

In Fig.~4 we present the internal energy of graphane as a function
of temperature, as derived from our PIMD simulations (solid circles).
As noted in Sec.~II.A, PIMD simulations yield separately
the potential ($E_{\rm pot}$) and kinetic ($E_{\rm kin}$) contributions 
to the internal energy $E$ \cite{he82,tu10,he14},
so that for $\tau = 0$ we have $E - E_0 = E_{\rm kin} + E_{\rm pot}$.
Solid squares and diamonds in Fig.~4 correspond to the kinetic and 
potential energy, respectively.
The internal energy $E - E_0$ is found to converge at low $T$ to 
479 meV/(C--H pair), which gives the zero-point energy of the system.
For comparison, we also display in Fig.~4 results of the internal 
energy obtained
in classical MD simulations (open circles). These data points lie
very close to the classical harmonic expectation, i.e.,
$E^{\rm cl} - E_0 = 6 k_B T$ per C--H pair. For $T \gtrsim 1000$~K we
observe a slight deviation of the simulation results from the 
harmonic expectancy, because of the onset of anharmonicity.
At high temperatures, the energy obtained from quantum simulations
converges to that of classical MD simulations. At $T = 1500$~K,
however, one still sees a significant difference between classical 
and quantum energy values.

For a purely harmonic model of the vibrational modes, one has
$E_{\rm kin} = E_{\rm pot}$ (virial theorem \cite{la80,fe72}),
irrespective of temperature in both classical and quantum approaches.
In our simulations of graphane, a ratio 
$E_{\rm kin} / E_{\rm pot} = 1$ is obtained for the classical model in
the low-temperature limit, as in this case the atomic motion does
not explore the energy landscape far from the absolute minimum,
because of the vanishingly small vibrational amplitudes.
This does not happen for the quantum results in the limit $T \to 0$,
since in this case the vibrational amplitudes remain finite, thus
feeling the anharmonicity of the interatomic potential.

In the quantum results we observe that $E_{\rm pot} > E_{\rm kin}$
in the whole temperature range shown in Fig.~4. The difference
$E_{\rm pot} - E_{\rm kin}$ increases for rising $T$, and remains
positive in the low-temperature limit. For $T \to 0$ we find a
difference of 8 meV/(C--H pair). This is basically
due to the zero-point expansion of graphane in both the
out-of-plane and in-plane directions (an anharmonic effect), 
which causes an increase of the potential energy with respect to 
the minimum energy configuration. At 1500 K the difference 
$E_{\rm pot} - E_{\rm kin}$ amounts to 24 meV per C--H pair,
i.e., three times larger than at low $T$.
The data presented in Fig.~4 correspond to $N = 96$.
For other cell sizes we obtained results for the internal energy
very close to those given for $N = 96$, and are
indistinguishable one from the other at the scale of the figure.

\begin{figure}
\vspace{-0.6cm}
\includegraphics[width=7cm]{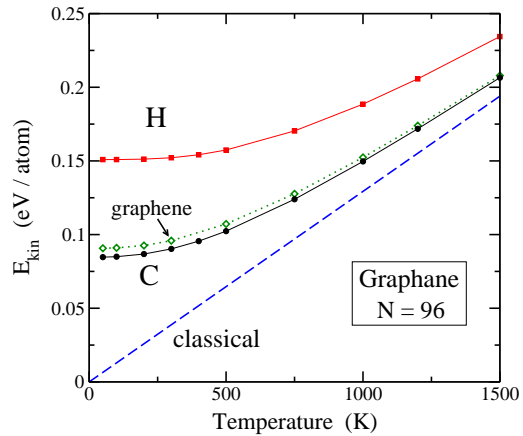}
\vspace{-0.5cm}
\caption{
Kinetic energy of hydrogen (solid squares) and carbon atoms
(solid circles) vs the temperature, as derived from PIMD simulations
of graphane. Open diamonds indicate the kinetic energy of carbon
atoms in graphene.
The dashed line represents the classical limit of the kinetic
energy per atom in a harmonic approximation:
$E_{\rm kin}^{\rm cl} = 3 k_B T / 2$.
Solid and dotted lines are guides to the eye.
}
\label{f5}
\end{figure}

The kinetic energy of hydrogen and carbon atoms is displayed
vs the temperature in Fig.~5. Solid symbols represent results of
PIMD simulations of graphane: squares for H and circles for C.
At low temperature, the kinetic energy for C is about one half of
that for H. This difference is reduced for increasing $T$, since 
the kinetic energy converges to the classical limit at high
temperature. The dashed line in Fig.~5 represents the classical
kinetic energy per atom: $E_{\rm kin}^{\rm cl} = 3 k_B T / 2$.
In this figure we have also plotted $E_{\rm kin}$ for C atoms in
graphene (open diamonds). At low $T$ it is somewhat larger
than that corresponding to C atoms in graphane (6 meV/atom).
The smaller value of $E_{\rm kin}$ in graphane is mainly due to 
a softening of vibrational modes such as C--C stretching
($sp^3$ hybridization in graphane vs $sp^2$ in graphene) \cite{ra19}.

A splitting of the potential energy of graphane into contributions
of hydrogen and carbon, similar to that presented for the kinetic 
energy, is not directly feasible from the results of PIMD simulations.
The virial estimator employed here to calculate $E_{\rm kin}$
(see Sec.~II.A) yields separately the contributions of H and C atoms.
However, the Hamiltonian corresponding to the TB method does not
allow to express independent inputs for the potential energy of
both species.  An indirect method to split $E_{\rm pot}$ could be
based on a separation into different vibrational modes of graphane,
but this would require to deal with a harmonic approximation for
the modes, as well as for the splitting of the energy of each mode
into the H and C parts.

A quantification of the overall anharmonicity in graphene 
can be found from the relation between $E_{\rm kin}$ and 
$E_{\rm pot}$. From the data presented above we find a ratio
$E_{\rm pot} / E_{\rm kin} =$~1.03 for low temperature, which
slowly increases for rising $T$, reaching a value of 1.06 at 1500~K.
Concerning the anharmonicity at low $T$, it is noteworthy that earlier 
analyses based on quasiharmonic approximations and perturbation theory 
indicate that the low-temperature changes in the vibrational energy with 
respect to a harmonic calculation are mostly due to the kinetic energy.
This is due to the fact that for a perturbed harmonic oscillator
at $T = 0$, the first-order change in the energy is given by
a variation of $E_{\rm kin}$, while $E_{\rm pot}$ is invariable
with respect to its unperturbed value \cite{la65,he95}.
In particular, this has been observed for the vibrational energy of
graphene at low temperature \cite{he16}.
In the results presented here for graphane we find
$E_{\rm pot} > E_{\rm kin}$, since the potential energy includes
an important anharmonic contribution due to changes in the layer
area, even at $T = 0$ with respect to the classical minimum
(see Sec.~VII).
The contribution of this {\em elastic} energy is not negligible,
even at low temperature, and can be obtained for a single layer
of graphene by reference to a strictly flat sheet, but it is not 
straightforwardly found for graphane which displays a finite
lateral dimension even for the classical minimum-energy 
configuration.

\section{Interatomic distances}

\begin{figure}
\vspace{-0.6cm}
\includegraphics[width=7cm]{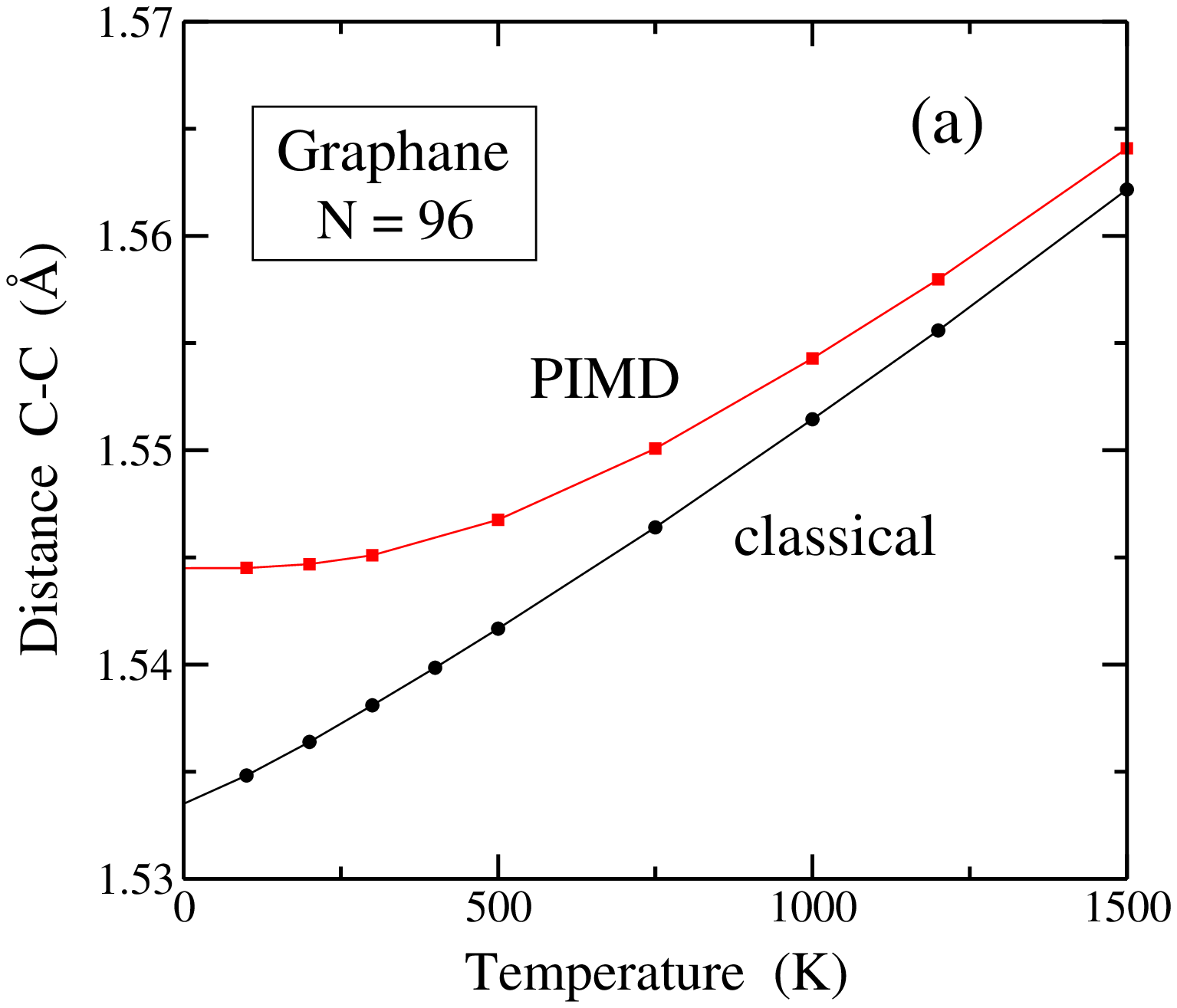}
\vspace{-0.0cm}
\includegraphics[width=7cm]{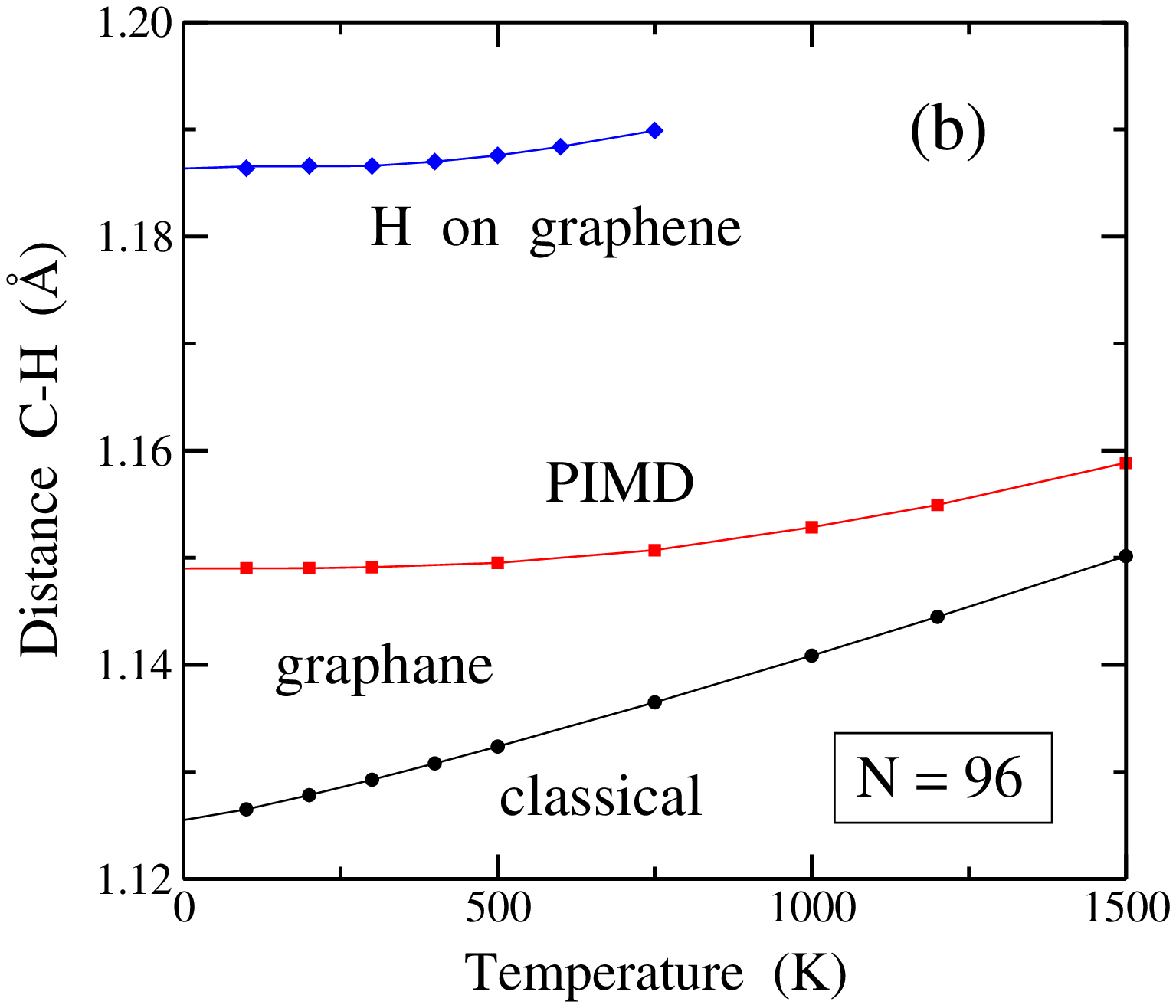}
\vspace{-0.5cm}
\caption{
Temperature dependence of mean interatomic distances in graphane.
(a) C--C distance; (b) C--H distance.
Solid circles and squares represent results of classical and
PIMD simulations, respectively.
Diamonds in (b) indicate the C--H distance for a single H atom on
graphene.  Lines are guides to the eye.
Error bars are less than the symbol size.
}
\label{f6}
\end{figure}

Here we present results for interatomic distances in graphane.
In Fig.~6(a) we show the temperature dependence of the equilibrium
C--C distance, $d_{\rm C-C}$, yielded by our PIMD simulations
(solid squares).
In the low-temperature limit $T \to 0$, we find an interatomic
distance of 1.5445 \AA, typical of a C--C single bond, 
which increases for rising temperature.
The size effect of the finite simulation cells on $d_{\rm C-C}$ is 
negligible. For $N$ = 96 and 216 we obtained differences similar to 
the error bars found for each cell size (less than the symbol size 
in Fig.~6(a)). We note that PIMD simulations
of graphene using the TB model employed here yielded a 
zero-temperature interatomic distance $d_{\rm C-C}$ = 1.4287 \AA,
intermediate between a single and a typical double bond
($\sim 1.34$ \AA).

For comparison with the interatomic distances derived from
the quantum simulations, we also display in Fig.~6(a) the temperature
dependence of $d_{\rm C-C}$, as obtained from classical MD simulations
(circles).  These data show at low temperature a nearly linear 
increase, as expected for interatomic distances and lattice parameters 
of crystalline solids in a classical approximation \cite{ki66,he00c}.
The classical results for $d_{\rm C-C}$ of graphane converge
at low temperature to an interatomic distance of 1.5337 \AA,
corresponding to the minimum energy configuration of graphane.
This value is close to the distance $d_{\rm C-C}$ given by
{\em ab-initio} calculations at $T = 0$ \cite{ca10}.
Our PIMD simulations predict at low $T$ a C--C distance larger than
the classical calculation, due to zero-point motion of the carbon
atoms along with anharmonicity of the interatomic potential.
Thus, for $T \to 0$ we find a zero-point expansion of the C--C bond
by $1.1 \times 10^{-2}$ \AA, which means an increase of a 0.7\%
with respect to the classical value.
We note that this rise in mean bond length due to nuclear quantum
effects is much larger than the precision reached in the determination
of interatomic distances from diffraction
techniques \cite{ya94,ra93b,ka98}.

The bond expansion caused by nuclear quantum effects is reduced for
rising temperature, as at high $T$ the quantum and classical predictions
should approach one to the other. Nevertheless, at 1500~K the
C--C distance derived from PIMD simulations is still
clearly larger than that found in the classical simulations.
The increase in $d_{\rm C-C}$ obtained in PIMD simulations
from $T = 0$ to room temperature is small, amounting to
$\sim 6 \times 10^{-4}$ \AA, i.e., about 27 times less
than the zero-point expansion.
Moreover, the zero-point bond expansion is similar
to the thermal expansion predicted by the classical model
from $T= 0$ to 650~K.

For comparison with the C--C distance in graphane, we note that
the TB potential employed here yields for a single layer of
graphene in the low-$T$ limit: $d_{\rm C-C}$ = 1.4192 \AA\ and
1.4265 \AA\, from classical and PIMD simulations, respectively.
This represents in the case of graphene a zero-point bond
expansion of a 0.5\% with respect to the classical prediction, 
somewhat less than for the relatively softer C--C bond in graphane.

In Fig.~6(b) we present the C--H bond distance in graphane
vs the temperature. As in Fig.~6(a) circles and squares are
data points obtained from classical MD and PIMD simulations,
respectively. The classical results converge at low $T$ to 
1.1257 \AA, a value close the result of {\em ab-initio} 
calculations \cite{sa10b,hu13b,ch14}, while
the low-temperature PIMD data yield $d_{\rm C-H}$ = 1.1490 \AA.
This means a zero-point expansion of 0.023 \AA, i.e., a 2\%
of the bond length. Note the larger relative increase in bond
distance compared with the C--C bond (a 0.7\%), due essentially
to the light mass of hydrogen.

We also show in Fig.~6(b) the distance C--H for a single
hydrogen impurity on graphene \cite{he09a}, as derived from PIMD
simulations, for comparison with the results for graphane.
For the single impurity, $d_{\rm C-H}$ is clearly larger than
in the case of graphane, converging to 1.1863 \AA\ for
$T \to 0$. This larger distance reflects a weakening of the
C--H bond with respect to the C--H bonds in graphane, where
carbon atoms display an $sp^3$ hybridization.
For a single hydrogen, however, the C atom adjacent to H
presents a local configuration intermediate between that required
by planar graphene ($sp^2$ hybridization) and tetrahedral
$sp^3$ adequate for the C--H bond.
In our simulations, a single hydrogen on graphene is found to
diffuse along the simulations at $T \sim 1000$~K, so that
a precise value for the C--H bond distance cannot be obtained
at these temperatures.

The general trend of classical and quantum data for $d_{\rm C-C}$ 
in a graphane layer is qualitatively similar to that found in 
simulations of 3D carbon-based materials such as diamond \cite{he00c},
as well as in graphene \cite{he16}.
The thermal bond expansion along with the zero-point dilation
presented here are an indication of anharmonicity in the interatomic
potential.  For C--C and C--H bonds in graphane, these effects are 
mainly due to anharmonicity in the corresponding stretching vibrations.
A more complex anharmonic effect emerges in the explanation 
of thermal variations in the in-plane area of 2D materials such
as graphane, because of the contributions of in-plane and
out-of-plane modes, as discussed below in Sec.~VII.

\section{Orientation of the C--H bonds}

To analyze the orientation of the C--H bonds we will 
consider spherical coordinates $(\theta,\phi)$.
The polar angle $\theta$ is defined as the angle between
the $z$-direction and the C--H bond, and $\phi$ is the azimuth
on the $(x,y)$ plane.
In the minimum-energy configuration we find $\theta = 0$, i.e.,
the C--H bond is strictly perpendicular to the $(x,y)$ plane.

\begin{figure}
\vspace{-0.6cm}
\includegraphics[width=7cm]{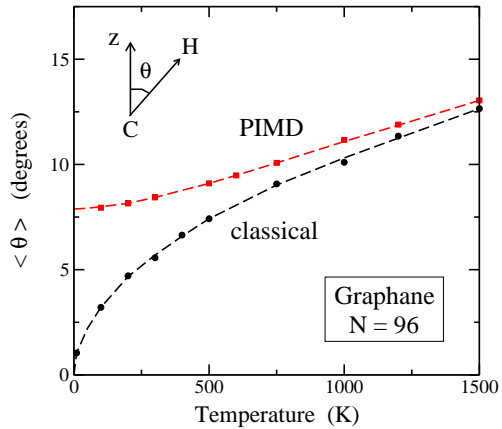}
\vspace{-0.5cm}
\caption{
Mean polar angle $\langle \theta \rangle$ between the
C-H bond and the $z$-direction as a function of
temperature. Symbols indicate results of simulations:
circles, classical MD; squares, PIMD.
Lines are guides to the eye.
Error bars are in the order of the symbol size.
}
\label{f7}
\end{figure}

In Fig.~7 we display the mean value of the polar angle, 
$\langle \theta \rangle$, for graphane as a function of temperature. 
Shown are results derived from classical (circles) and PIMD 
(squares) simulations.
In the classical approach we find that $\langle \theta \rangle$
converges to zero at low temperature as $\sqrt{T}$.
In fact, we find a dependence $\langle \theta \rangle^2 = \gamma T$
with a coefficient $\gamma$ = 0.110 deg$^2$/K.
The PIMD results converge to a finite value for $T \to 0$:
$\langle \theta \rangle_0$ = 7.8 deg due to zero-point motion.
$\langle \theta \rangle$ increases as temperature is raised 
and reaches a value
of 13.0 deg for $T$ = 1500 K. At this temperature the classical 
and quantum results are close one to the other.

Our results for $\langle \theta \rangle$ can be related with
the probability distribution for the direction of the C--H bond
in the whole sphere.
We describe the orientation of the C--H axis by the probability 
density $P(\theta, \phi)$, which verifies the normalization
condition:
\begin{equation}
   \int_0^{2 \pi} d \phi  \int_0^{\pi}
           P(\theta,\phi) \sin \theta \, d \theta  = 1    \, .
\end{equation}
Although $P(\theta, \phi)$ could depend on the azimuthal angle $\phi$, 
we find a uniform distribution for $\phi \in [0, 2\pi]$, 
i.e. we observe a density with cylindrical symmetry around 
the $z$ axis (changes in the probability density for different 
angles $\theta$ are less than 3\%).
Thus, we can define an average density 
\begin{equation}
  \bar{P}(\theta) =  \int_0^{2 \pi}  P(\theta,\phi) \, d \phi  \, ,
\end{equation}
which depends only on the polar angle $\theta$.

The results of our simulations, both classical and quantum,
indicate that $\bar{P}(\theta)$ follows very closely (i.e., within
the statistical noise) a Gaussian distribution:
\begin{equation}
    \bar{P}(\theta) = c \, \exp(- a \theta^2) \, ,
\label{ptheta}
\end{equation}
where $c$ is a normalization constant given by
\begin{equation}
  c^{-1} = \int_0^{\pi} \sin \theta \, \exp(- a \theta^2) \, d \theta  \, .
\end{equation}
The parameter $a$ in Eq.~(\ref{ptheta}) controls the width of the
Gaussian distribution.  It decreases for increasing $T$ 
as the density distribution becomes wider.
From the results of our PIMD simulations, we find
$a$ = 40.6 and 14.8 rad$^{-2}$ for $T$ = 100 and 1500 K,
respectively.

In the classical approach, the parameter $a$ diverges for $T \to 0$,
and the Gaussian in Eq.~(\ref{ptheta}) converges to a Dirac
$\delta$-function. Then, the C--H bonds are strictly perpendicular
to the $(x,y)$ layer plane, as indicated above.  In general, 
the mean polar angle $\langle \theta \rangle$ can be written as
\begin{equation}
  \langle \theta \rangle =   \int_0^{\pi} 
        \bar{P}(\theta) \, \theta \,  \sin \theta \, d \theta  \; .
\end{equation}
In the quantum approach we find a zero-temperature limit
$\langle \theta \rangle_0$ = 0.136 rad = 7.8 deg, which corresponds
to a parameter $a_0$ = 42.2 rad$^{-2}$.

\section{Atomic motion}

In this section we present results for the MSD
of C and H atoms in graphane.
We concentrate on the nature of the atomic displacements,
to find out if they may be described by classical motion or the atoms
largely behave as quantum particles. 
We expect that a quantum model will be more appropriate at low temperature,
not only for H, but also for the relatively heavier C atoms.
We employ the notation presented in Sec.~II.C.

\begin{figure}
\vspace{-0.6cm}
\includegraphics[width=7cm]{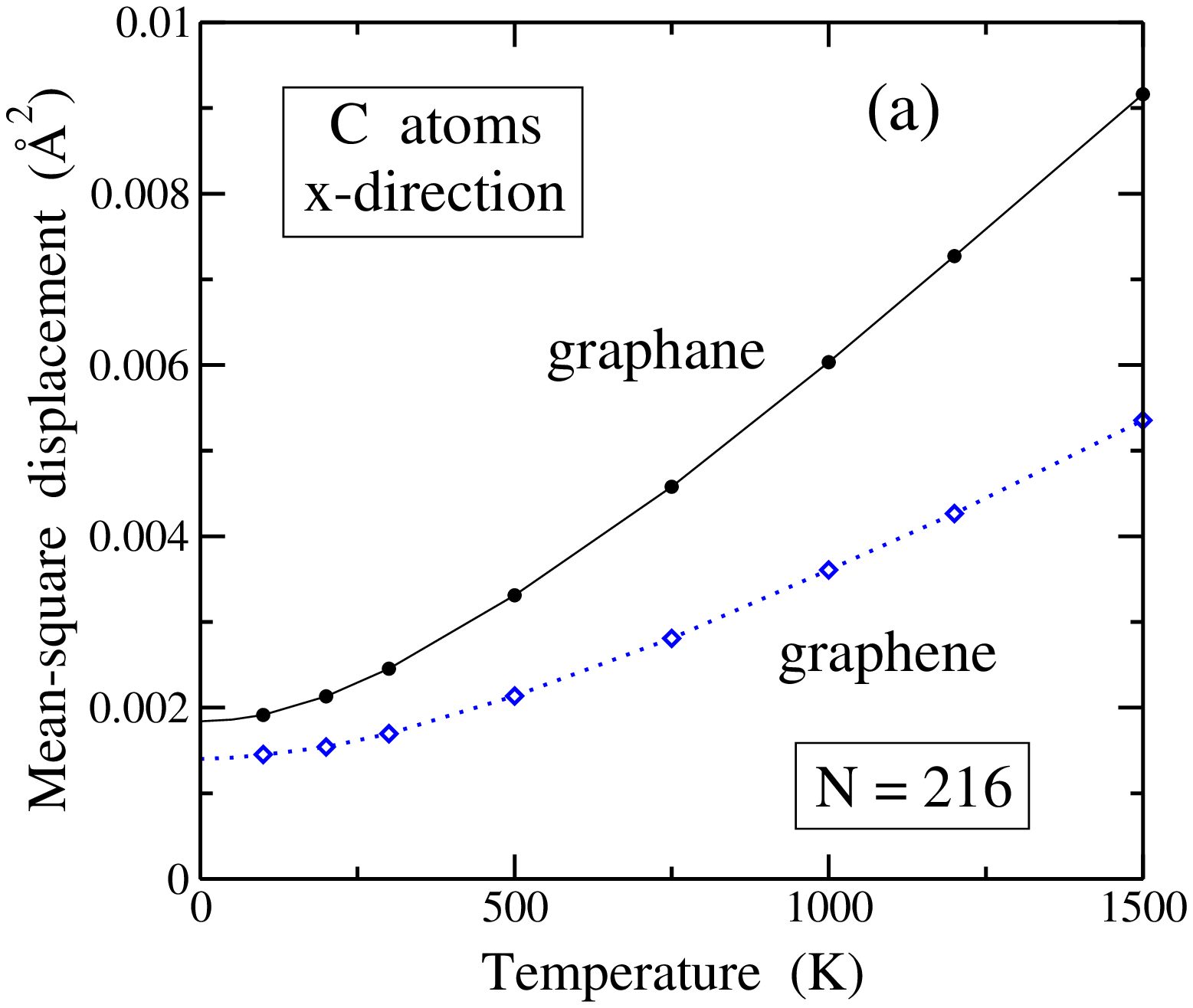}
\vspace{-0.0cm}
\includegraphics[width=7cm]{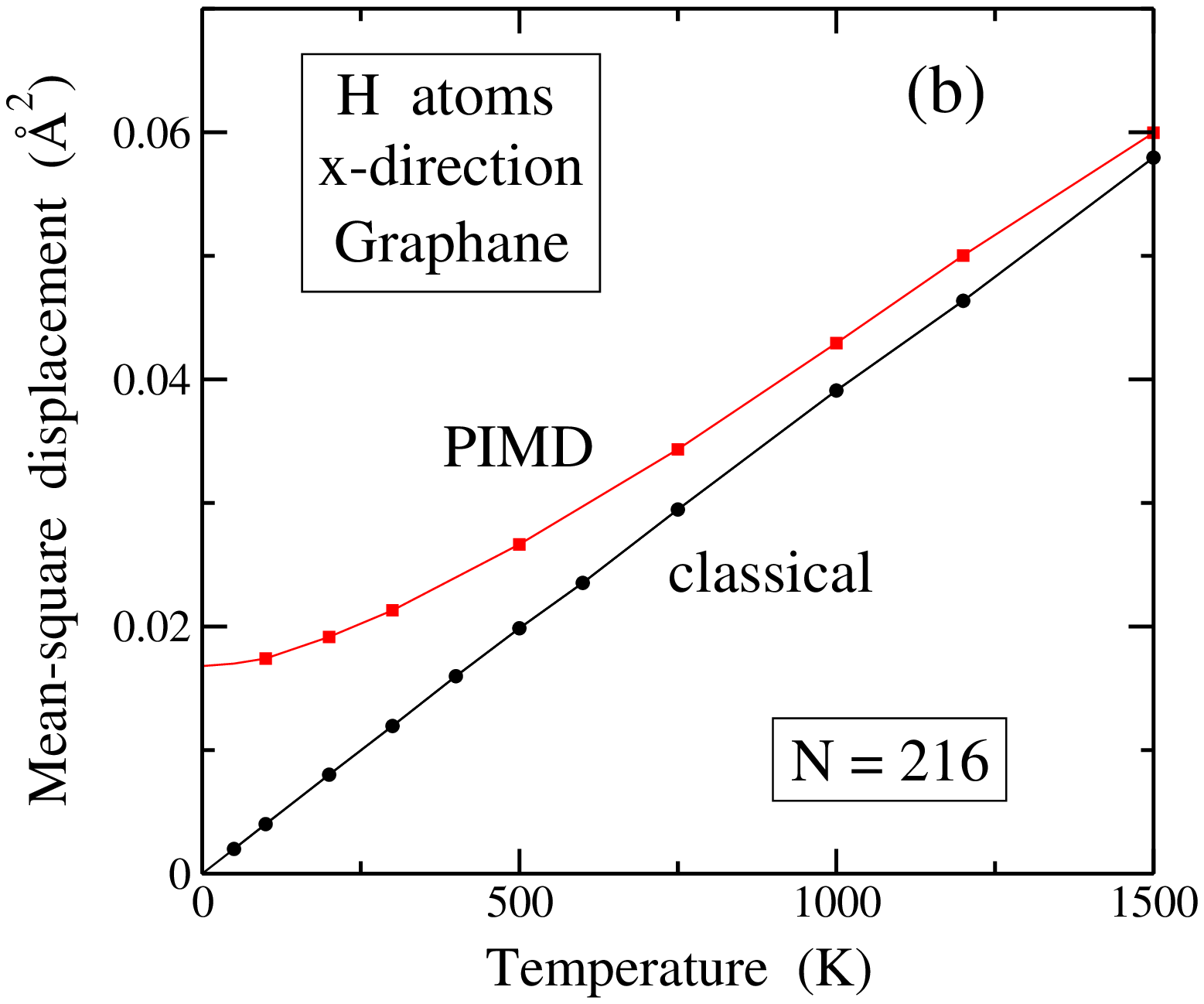}
\vspace{-0.5cm}
\caption{
Atomic mean-square displacements in the in-plane $x$-direction.
(a) Carbon atoms: Symbols are data points obtained from PIMD
simulations for graphane (solid circles) and graphene
(open diamonds).
(b) Hydrogen atoms: Symbols are data points derived from
classical (circles) and PIMD simulations (squares).
These results were obtained for a graphane cell containing
216 C-H pairs.
Lines are guides to the eye.
}
\label{f8}
\end{figure}

In Fig.~8 we show the MSD of C and H atoms on the $(x,y)$ plane.
Results for the $x$ and $y$ directions turn out to be indistinguishable, 
so we present only those found in the $x$-direction.
These data were obtained for a simulation cell with $N$ = 216.
In Fig.~8(a) we display the MSD of C atoms as derived from 
PIMD simulations of graphane (solid circles).
For comparison we also present the MSD of carbon atoms in graphene,
obtained by using the same TB model.
The MSD is clearly larger in the case of graphane,
reflecting a softer effective in-plane potential for the motion
of C atoms in the $(x,y)$ plane. In other words, this is due to
a reduction in the vibrational frequencies of in-plane acoustic modes 
LA and TA in graphane with respect to graphene \cite{ra19},
which causes larger amplitudes in the former case.

In Fig.~8(b) we display the mean-square displacement
$(\Delta x)_{\rm H}^2$ of hydrogen atoms in graphane as a function 
of temperature.
Shown are data points obtained from classical (circles) and
PIMD (squares) simulations.
The quantum results converge at low $T$ to a value 
$(\Delta x)_0^2 = 0.017$ \AA$^2$.
Comparing the room-temperature data ($T = 300$~K), we find that
the quantum result is 1.8 times larger than the classical one.
This ratio increases as the temperature is lowered, and at 100 K
it amounts to 4.3.
Comparing the quantum results for the MSD of C and H atoms in graphane 
at room temperature, we find that the former is about 9 times smaller
than the latter.

\begin{figure}
\vspace{-0.6cm}
\includegraphics[width=7cm]{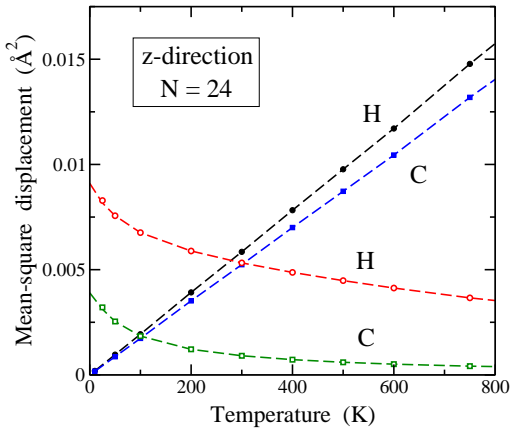}
\vspace{-0.5cm}
\caption{
Temperature dependence of the mean-square displacements along
the out-of-plane direction:
classical $C_z^2$ (solid symbols) and quantum $Q_z^2$ (open symbols).
In both cases, circles correspond to hydrogen and squares to carbon
atoms.  These results were obtained for a graphane cell containing
24 C-H pairs.  Lines are guides to the eye.
}
\label{f9}
\end{figure}

We now turn to the out-of-plane motion.
This motion is relevant for several properties of 2D materials, as
it controls the appearance of bending and crumpling in their
constituent layers.
In Fig.~9 we present results for the MSD of H and C in the
$z$-direction, obtained for a graphane cell with $N = 24$.
The reason for presenting here results corresponding to a cell size
smaller than in previous figures is that for $N = 24$ one can visualize
more clearly the competition between quantum and classical 
contributions to the MSD.
As explained in Sec.~II.C, this displacement may be divided 
into two parts: one of them quantum in nature, which
corresponds to the spread of the quantum paths, $Q_z^2$ (open symbols),
and another of classical character, which accounts for motion of
the centroid (global displacements of the paths), $C_z^2$ 
(solid symbols).
For $T \to 0$, $C_z^2$ vanishes and $Q_z^2$ converges to zero-point
values of 3.9 and $9.1 \times 10^{-3}$ \AA$^2$ for C and H, 
respectively. $Q_z^2$ decreases for rising $T$, as the spatial
extension of the quantum paths decreases, while $C_z^2$ grows
almost linearly, in accord with the expectancy for the MSD of
classical particles. 

For the system size shown in Fig.~9 (N = 24),
both terms contributing to $(\Delta z)^2$ are nearly equal at 
$T$ = 105~K and 280~K for carbon and hydrogen, respectively.
For each element, at higher temperatures 
the classical contribution $C_z^2$ dominates
the atomic displacements on the $z$-direction.
The actual quantum delocalization of the atoms in the out-of-plane
direction can be estimated from the ``mean extension,''
$(\Delta z)_Q$, of the quantum paths in this direction.
At 300~K our simulations yield an average extension 
$(\Delta z)_Q = (Q_z^2)^{1/2}$ = 0.030 and 0.073 \AA, for carbon
and hydrogen, respectively. This mean extension increases with
lowering temperature, and for $T \to 0$, $(\Delta z)_Q$ amounts
to 0.062 \AA\ for C and 0.095 \AA\ for H atoms.

At finite temperatures, the MSD in the out-of-plane direction,
$(\Delta z)^2$, increases much faster than $(\Delta x)^2$, as a
consequence of the rise in the classical contribution $C_z^2$ in the
$z$ direction.
This is due to the presence of long-wavelength vibrational modes
with low frequency and large vibrational amplitudes in the
ZA {\em flexural} band (atomic displacements in the $z$-direction).
Calling ${\bf k} = (k_x, k_y)$ the wavevectors in the 2D reciprocal 
lattice of graphane \cite{ra19},
this phonon band can be described at finite temperatures by
a dispersion relation of the form
$\rho \, \omega({\bf k})^2 = \sigma k^2 + \kappa k^4$, where
$k = |{\bf k}|$, $\rho$ is the surface mass density,
$\sigma$ an effective stress,
and $\kappa$ the so-called {\em bending modulus} \cite{ra16}.
For our present purposes, $\sigma$ is negligible and the
flexural band may be considered as parabolic:
$\omega({\bf k}) \approx \sqrt{\kappa/\rho} \, k^2$.
For the present TB model we find at $T =$ 300~K a bending modulus
$\kappa$ = 1.4 eV \cite{ra19}.
For increasing system size $N$ there appear vibrational modes
with longer wavelength $\lambda$. In practice, one has an effective
cut-off $\lambda_{max} \approx L$, where $L = (N A_p)^{1/2}$,
and $A_p$ is the in-plane area per C atom (see below).
Thus, we have
$k_{min} = 2 \pi / \lambda_{max}$, which means
$k_{min} \sim N^{-1/2}$.

For system sizes larger than $N$ = 24,
the temperature dependence of the atomic MSD in the $z$ direction 
is similar to that shown in Fig.~9.
The main difference is that the temperature range
where $Q_z^2$ or $C_z^2$ is the dominant contribution to 
$(\Delta z)^2$ depends on $N$. 
This is caused by the enlargement of the classical part $C_z^2$
for rising size, while $Q_z^2$ is rather insensitive to $N$
(clear finite-size effects in $Q_z^2$ are only found for very
small simulation cells).  
This is analogous to earlier observations in graphene \cite{he16}.
For given system size $N$ and atomic species, the ratio 
$Q_z^2 / C_z^2$ decreases for rising $T$, so that there appears 
a crossover temperature $T_c$ for which this ratio equals unity,
as indicated above for $N = 24$. 
The main difference with graphene is that in the case of graphane
each species sets its own temperature scale (or $T_c$) 
for this purpose.
In each case, for $T > T_c$ classical-like motion is the dominant 
contribution  in the atomic MSD in the $z$ direction.
For graphene, the temperature $T_c$ was found to decrease for 
increasing system size $N$ as a power-law \cite{he16}, such as
$T_c \sim N^{-b}$ with an exponent $b = 0.67$.
With the present TB model, however, we cannot reach large system
sizes to obtain a reliable value for the exponent $b$ in graphane.

The competition between $Q_z^2$ and $C_z^2$ as functions
of the system size presented here does not emerge for
the in-plane MSD, where the crossover temperature
$T_c$ is rather insensitive to the system size.
For motion in the $(x, y)$ plane, $Q_x^2$ and $C_x^2$ quickly 
converge for rising $N$ to their corresponding asymptotic limit, 
in a way similar to the MSD derived from vibrational motion 
in 3D solids \cite{he14}.
The main difference between in-plane and out-of-plane 
vibrational motion in this context is the appearance of the
flexural ZA band in the $z$-direction with its distinctive
parabolic dispersion relation for small $k$,
$\omega_{\rm ZA} \sim k^2$, different from usual
acoustic modes with $\omega \sim k$ \cite{ra19}.

A consistency check for the overall results of our quantum
atomistic simulations of graphane can be found from 
the comparison of data
corresponding to the coordinate and momentum space. 
This can be done through the MSDs in real space and 
the kinetic energy presented in Sec.~III.
In fact, according to Heisenberg's uncertainty principle,
$E_{\rm kin}$ should verify the relation
$E_{\rm kin} \geq F$, with the function $F$:
\begin{equation}
 F = \frac{\hbar^2}{8 m}  \left[ 
  (\Delta x)^{-2} + (\Delta y)^{-2} + (\Delta z)^{-2} \right] \, ,
\end{equation}
where $m$ is the particle mass (see the Appendix).

For a 3D harmonic oscillator, the ratio
$E_{\rm kin} / F$ converges to unity for $T \to 0$ (ground state).
In general, one can consider the function $F$ as a lower
boundary for the kinetic energy of a quantum particle.
From our PIMD simulations of graphane at 50 K, we find $F$ = 125 
and 57 meV for H and C, respectively. These values are smaller than
the corresponding kinetic energy per atom: $E_{\rm kin}$ =
149 and 86 meV, which give $E_{\rm kin} / F$ = 1.2 for hydrogen
and 1.5 for carbon.

The ratio $E_{\rm kin} / F$ converges at low $T$ to a value 
higher than 1 when one has a frequency dispersion for the
vibrational modes, as is usual in condensed matter.
Thus, for the well-known Debye model of solids \cite{ki66}, 
$E_{\rm kin} / F \to 1.125$ for $T \to 0$ (see the Appendix).
For hydrogen in graphane we find at low-$T$ a ratio of 1.2, 
due to frequency dispersion and anisotropy of this material 
(in-plane vs out-of-plane modes), which causes an increase in
$E_{\rm kin} / F$.  For C atoms we obtain a higher value at 50~K, 
which indicates that in this case we are farther from the
low-$T$ limit due to the larger atomic mass.

\section{Layer area}

The simulations (both classical MD and PIMD) presented here were carried
out in the isothermal-isobaric ensemble, as explained in Sec.~II.A.
This means that in a simulation run we fix the number $N$ of C--H pairs, 
the temperature $T$, and the applied stress in 
the $(x, y)$ plane ($\tau = 0$ in our simulations),
thus allowing for changes in the area of the simulation cell
on which periodic boundary conditions are applied.

\begin{figure}
\vspace{-0.6cm}
\includegraphics[width=7cm]{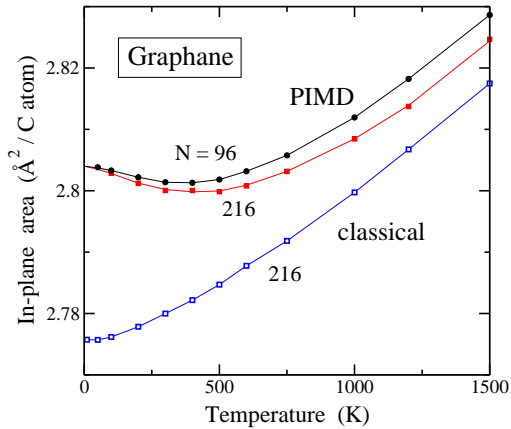}
\vspace{-0.5cm}
\caption{
Temperature dependence of the mean in-plane area $A_p$ of graphane.
Solid symbols correspond to results of PIMD simulations for
$N$ = 96 (circles) and 216 (squares).
Open squares are data points derived from classical MD simulations
for $N = 216$.
Error bars are less than the symbol size.
Lines are guides to the eye.
}
\label{f10}
\end{figure}

In Fig.~10 we present the temperature dependence of the in-plane area
per C atom, $A_p = L_x L_y / N$, as derived from classical MD 
(open squares) and quantum PIMD simulations (solid squares) of 
graphane for $N = 216$.  For comparison we also display results from 
PIMD simulations for $N = 96$.
In the results of classical simulations we observe a slight decrease 
in $A_p$ at low $T$ (almost unobservable at the scale of Fig.~10), 
and at higher temperatures the in-plane area 
increases for rising $T$.
In the low-$T$ limit, the classical in-plane area converges to the
minimum-energy configuration with $A_0 = 2.7758$~\AA$^2$.
The general trend of these data is similar to that found in earlier 
classical Monte Carlo and MD simulations of graphene 
single layers \cite{za09,ga14,br15}.
In the results of PIMD simulations we observe a clear decrease in
$A_p$ from the low-temperature limit to $T \approx 400$~K, and a
rise of $A_p$ at higher temperature.

The main difference between $A_p$ for different system sizes is
a decrease in the in-plane area for rising cell size, but the
temperature of the minimum $A_p$ is roughly unaffected.
Notwithstanding the differences in the in-plane area 
per atom for the different system sizes, all of them converge 
in each case (quantum or classical) to a single value 
at low temperature.
For $T \to 0$, the difference between $A_p$ derived from quantum
and classical simulations amounts to 0.029 \AA$^2$/(C atom). 
This difference decreases as temperature is raised,
since nuclear quantum effects become less important.
At 1500 K it is $7 \times 10^{-3}$ \AA$^2$/(C atom).
Note that $d A_p / d T$ has to vanish for $T \to 0$, as required by
the third law of Thermodynamics \cite{ca60}. This is the case 
for the results of PIMD simulations for graphane, and has been also 
discussed earlier for graphene \cite{he16}.

The behavior of $A_p$ as a function of temperature can be explained
as due to two competing factors.
On one side, there is a tendency of the C--C distance to increase
for rising $T$ (see Fig.~6(a)), which favors a rise in $A_p$.
On the other side, bending of the whole graphane layer
gives rise to a reduction of its projection on the $(x, y)$ plane,
i.e. the area $A_p$.
For low temperature, the increase due to the first factor is
dominated by the second one (bending), so that $d A_p / d T < 0$.
This is especially appreciable in the quantum results, since in this
case the thermal expansion of the C--C bonds at low $T$ is small.
At high temperatures, the rise in C--C distance dominates the reduction
in the in-plane area due to out-of-plane atomic displacements,
and we have $d A_p / d T > 0$.

In connection with changes in the area $A_p$, we define 
the in-plane thermal expansion coefficient as
\begin{equation}
 \alpha_p = \frac{1}{A_p}
      \left( \frac{\partial A_p}{\partial T} \right)_{\tau}   \, .
\end{equation}
which is negative at low $T$ and positive at high $T$, according
to our PIMD simulations of graphane. $\alpha_p$ vanishes when
the curve $A_p(T)$ has a minimum, i.e., for $T \approx 400$~K.
At room temperature ($T = 300$~K) we find 
$\alpha_p = -2.6 \times 10^{-6}$ K$^{-1}$.
Note, in contrast, that classical simulations yield at this 
temperature a positive coefficient 
$\alpha_p = 7.8 \times 10^{-6}$ K$^{-1}$.

For a single graphene layer, PIMD simulations with the TB model 
employed here yield a temperature dependence of the in-plane area 
analogous to that found earlier using effective interatomic 
potentials, such as the so-called LCBOPII 
(a long-range carbon bond-order potential) \cite{he16}.
For graphene, the minimum area is found at $T \sim 1200$~K,
a temperature much larger than that obtained for graphane.
This important difference is due to the fact that the thermal
expansion of the C--C bond in graphane is larger than that
in graphene.  Thus, 
at $T$ = 500 K we find an increase in $d_{\rm C-C}$ of
$1.1 \times 10^{-5}$~\AA/K for graphane vs 
$4.4 \times 10^{-6}$~\AA/K for graphene, i.e,
the former thermal expansion is a factor of 2.5 larger than
the latter. At 1000 K we find a ratio of 2.3.

As a brief summary of the data presented in this section, 
we point out that changes in the in-plane area are important anharmonic 
effects, to which the PIMD procedure is very sensitive. 
At low temperatures, these anharmonic effects are appreciably
enlarged by quantum motion, as shown in Fig.~10. 
This is caused by the fact that anharmonicity is revealed
in the quantum model even at low $T$, in contrast to the classical
case, where it progressively appears for rising temperature.
We have also found that size effects in the in-plane area $A_p$ are 
much less important than in graphene.

\section{Compressibility}

Important physical information about the intrinsic stability of 2D
materails can be obtained by studying their mechanical properties.
In particular, properties such as stiffness and bending rigidity
can be affected by crumpling or corrugation of 
the layers \cite{ru11,ko13,ko14},
and are relevant for possible applications \cite{se10,pr10}.
An interesting variable in this context is the 2D compressibility,
which can be directly calculated from PIMD simulations.

The in-plane isothermal compressibility, $\chi_p$, at temperature $T$,
is defined as
\begin{equation}
   \chi_p = - \frac{1}{A_p} 
            \left( \frac{\partial A_p}{\partial \tau} \right)_T   \, .
\label{chip1}
\end{equation}
In this equation, the variables in the r.h.s. correspond to in-plane 
variables, since the stress $\tau$ in the isothermal-isobaric
ensemble employed here is a variable conjugate to the in-plane 
area $A_p$.    The inverse of $\chi_p$, $B_p = 1 / \chi_p$,
is the 2D modulus of hydrostatic compression \cite{be96b}, with
units of eV/\AA$^2$ or N/m.
$B_p$ is the in-plane analogous to the bulk modulus of 3D solids. 

An alternative way to calculate $\chi_p$ consists in using
the fluctuation formula \cite{la80,ra17}
\begin{equation}
   \chi_p = \frac{N \sigma_p^2}{k_B T A_p}
\label{chip2}
\end{equation}
where $\sigma_p^2$ are the mean-square fluctuations of the
area $A_p$ obtained in the simulations.
This expression turns out to be more adequate for our present purposes 
than obtaining $(\partial A_p / \partial \tau)_T$ as in 
Eq.~(\ref{chip1}), because 
a calculation of this derivative by numerical methods needs 
to carry out additional simulations at nonzero stresses.
Thus, using Eq.~(\ref{chip2}) we can calculate the compressibility
$\chi_p$ from our PIMD simulations with vanishing external 
stress ($\tau = 0$). In any case,
at some selected temperatures we have checked that
both methods yield the same results for $\chi_p$,
taking into account the statistical error bars.

\begin{figure}
\vspace{-0.6cm}
\includegraphics[width=7cm]{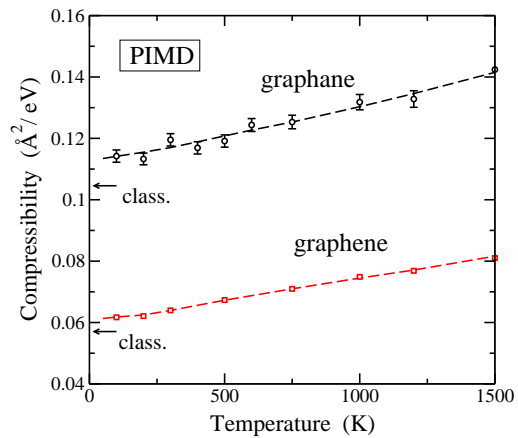}
\vspace{-0.5cm}
\caption{Temperature dependence of the in-plane compressibility
$\chi_p$, as obtained from PIMD simulations for graphane (circles)
and graphene (squares). Dashed lines are guides to the eye.
Error bars for the graphene data are in the order of the symbol
size. Horizontal arrows with the label ``class'' indicate the
zero-temperature classical limit for $\chi_p$.
}
\label{f11}
\end{figure}

In Fig.~11 we show the temperature dependence of the compressibility 
$\chi_p$ for graphane (circles), as derived from our
stress-free PIMD simulations. 
For comparison we also present results for graphene obtained with
the same method (squares).
Error bars for $\chi_p$ of graphane are about two or three times larger 
than those for graphene, as a consequence of the larger values of
$\sigma_p^2$ for the latter, i.e., fluctuations in $\chi_p$ rise
when its mean value increases.

The classical value of the in-plane compressibility at $T = 0$ 
is given by
\begin{equation}
    \chi_0 =  \frac{1}{A_0}  \left( 
         \frac{\partial^2 E}{\partial A_p^2}  \right)_0^{-1}  \; ,
\end{equation}
where the subscript ``0'' indicates that the derivative is taken
at the minimum-energy configuration (area $A_0$).
Horizontal arrows in Fig.~11 indicate the classical zero-temperature
limit $\chi_0$ for graphane and graphene.
We find that the compressibility appreciably rises when nuclear 
quantum effects are taken into account. In the low-temperature limit,
$\chi_p$ of graphane increases by a 9\% with respect to the
classical prediction, a relative growth
similar to that found for graphene.

\section{Summary}

Graphane, as a quasi-2D material, displays typical properties 
of membranes, but a detailed study at the atomic scale yields
important information about the effect of vibrational modes on
the properties of this material. 
We have presented and discussed results of PIMD simulations 
of graphane in the isothermal-isobaric ensemble in a wide
range of temperatures.
This kind of simulations have revealed themselves as an adequate
technique to study several properties of graphane layers.
We have focused on the importance of nuclear quantum effects,
which have been quantified by comparing results of the quantum
simulations with those given by classical MD simulations.
Our results indicate that
explicit consideration of the quantum nature
of atomic nuclei is important for a reliable description of 
these quasi-2D membranes, especially at low temperatures. 
Even for $T$ close to room temperature, such nuclear quantum effects
are not negligible.

Several structural variables have been found to change when
the quantum nature of atomic nuclei is taken into account.
The interatomic distances and in-plane area 
appreciably increase with respect to the classical prediction.
In the low-$T$ limit the mean C--C and C--H bonds increase by
0.7\% and 2\%, respectively. The higher value for
the C-H bonds is indeed due to the lighter mass of hydrogen,
combined with the anharmonicity of the interatomic potential.

Zero-point expansion of the graphane layer due to nuclear quantum 
motion amounts to about 1\% for the in-plane area $A_p$. 
For $T > 0$, a thermal contraction of $A_p$ is found for
graphane in a similar way to graphene monolayers.
However, it is important to note that in the case of graphane 
this contraction is almost unobservable in classical
MD simulations.
Thus, the temperature dependence of the area $A_p$ is qualitatively 
different when derived from classical or PIMD simulations
at temperatures around 300~K and even higher (see Fig.~10).
In the quntum simulations,
the thermal expansion coefficient $\alpha_p$ is found to
be negative for $T \lesssim$ 400~K,
and it becomes positive at higher $T$.

The in-plane compressibility of graphane $\chi_p$ has been found 
to be about twice that of graphene, as a consequence of the tetrahedral
(nonplanar) connectivity of the C atoms in graphane.
$\chi_p$ appreciably rises when nuclear quantum motion is
considered, and in the limit $T \to 0$, it is a 9\% larger than 
the classical expectancy.

Atomic vibrations in the out-of-plane direction increase with both
temperature and system size.  Although quantum effects show up in 
these vibrational modes, at finite temperatures classical-like motion 
overshadows the quantum delocalization in the $z$-direction, 
as long as the system size is large enough.
This size effect is observable in PIMD simulations at low temperatures,
due to the onset of vibrational modes with smaller wavenumbers
in larger cells, in particular in the ZA flexural band.

A quantitative assessment of the anharmonicity in the atomic
vibrations is given by a comparison of the kinetic and potential
energy of the system, which should coincide for strictly
harmonic modes.
For a graphane layer we find that
$E_{\rm pot} > E_{\rm kin}$ in the whole temperature
range studied here. The difference $E_{\rm pot} - E_{\rm kin}$
increases as $T$ is raised, and in the low-temperature limit
it amounts to 8 meV/(C--H pair), i.e., a 3.5\% of $E_{\rm kin}$.
This is caused by the anharmonicity of the system at low $T$, 
as probed by the atomic zero-point motion.

PIMD simulations similar to those presented here can yield
information on the structural and mechanical properties of 
graphane under tensile and compressive stress.
This may give us insight into the relative stability of these
layers in a stress-temperature phase diagram.  \\  \\

{\bf CRediT author contribution statement}  \\

Carlos P. Herrero: Data curation, Investigation, Validation, Original draft

Rafael Ram\'irez: Methodology, Software, Investigation, Validation  \\  \\

{\bf Declaration of Competing Interest}  \\

The authors declare that they have no known competing financial
interests or personal relationships that could have appeared to 
influence the work reported in this paper.  \\  \\

\begin{acknowledgments}
This work was supported by Direcci\'on General de Investigaci\'on,
MINECO (Spain) through Grants FIS2015-64222-C2 and PGC2018-096955-B-C44.
\end{acknowledgments}


\appendix

\section{Uncertainty relations}

The RMS displacements of
the position coordinate $x$ and momentum $p_x$ verify the Heisenberg's
uncertainty relation (see, e.g., Ref.~\cite{co77} complement C.III)
\begin{equation}
     \Delta x \, \Delta p_x \geq \frac{\hbar}{2}  \, ,
\label{xpx}
\end{equation}
and similar expressions are valid for the $y$ and $z$ coordinates.
Taking into account that  $\langle p_x \rangle = 0$, then
$(\Delta p_x)^2 = \langle p_x^2 \rangle$,
so that for a particle with mass $m$ we have:
\begin{equation}
 E_{\rm kin} = \frac{\langle {\bf p}^2 \rangle}{2 m} =  \frac{1}{2 m}  
   \left[ (\Delta p_x)^2 + (\Delta p_y)^2 + (\Delta p_z)^2 \right] \, .
\end{equation}
From the inequality in Eq.~(\ref{xpx}) it follows
\begin{equation}
   (\Delta p_x)^2  \geq  \frac{\hbar^2}{4 (\Delta x)^2}  \, ,
\end{equation}
which yields
\begin{equation}
    E_{\rm kin}  \geq  F   \, ,
\end{equation}
$F$ being a function of the atomic MSDs:
\begin{equation}
    F =  \frac{\hbar^2}{8 m} \left[ 
       (\Delta x)^{-2} + (\Delta y)^{-2} + (\Delta z)^{-2} 
         \right] \, .
\end{equation}
This means that we have a lower boundary for the kinetic energy
from the real-space delocalization.

For an isotropic 3D harmonic oscillator wiht frequency $\omega$,
we have for the ground state:
\begin{equation}
     (\Delta x)^2_0 =  \frac{\hbar}{2 m \, \omega} \, ,
\end{equation}
and
\begin{equation}
     (E_{\rm kin})_0 = \frac34 \hbar \, \omega \, ,
\end{equation}
so that $E_{\rm kin} / F$ converges to unity in the 
low-temperature limit.

To take into account the dispersion of frequencies in condensed
matter, we can consider an isotropic 3D Debye model \cite{ki66} 
with a vibrational density of
states $\mu(\omega) \propto \omega^2$ and a high-frequency
cutoff $\omega_D$. In this case, assuming harmonic vibrations,
we find for $T \to 0$:
\begin{equation}
 (\Delta x)^2_0 = \frac{3}{\omega_D^3} 
   \int_0^{\omega_D}  \frac{\hbar}{2 m \, \omega} \, \omega^2  d \omega
          =  \frac34 \, \frac{\hbar}{m \, \omega_D} \, ,
\end{equation}
and
\begin{equation}
     (E_{\rm kin})_0 =  3 \, \frac{(\Delta p_x)^2_0}{2m} =
             \frac{9}{16} \hbar \, \omega_D \, .
\end{equation}
Hence, in the low-$T$ limit the ratio $E_{\rm kin} / F$ for 
the Debye model is independent of $\omega_D$ and  
converges to 1.125.




\begin{thebibliography}{78}
\expandafter\ifx\csname natexlab\endcsname\relax\def\natexlab#1{#1}\fi
\expandafter\ifx\csname bibnamefont\endcsname\relax
  \def\bibnamefont#1{#1}\fi
\expandafter\ifx\csname bibfnamefont\endcsname\relax
  \def\bibfnamefont#1{#1}\fi
\expandafter\ifx\csname citenamefont\endcsname\relax
  \def\citenamefont#1{#1}\fi
\expandafter\ifx\csname url\endcsname\relax
  \def\url#1{\texttt{#1}}\fi
\expandafter\ifx\csname urlprefix\endcsname\relax\def\urlprefix{URL }\fi
\providecommand{\bibinfo}[2]{#2}
\providecommand{\eprint}[2][]{\url{#2}}

\bibitem[{\citenamefont{Geim and Novoselov}(2007)}]{ge07}
\bibinfo{author}{\bibfnamefont{A.~K.} \bibnamefont{Geim}} \bibnamefont{and}
  \bibinfo{author}{\bibfnamefont{K.~S.} \bibnamefont{Novoselov}},
  \bibinfo{journal}{Nature Mater.} \textbf{\bibinfo{volume}{6}},
  \bibinfo{pages}{183} (\bibinfo{year}{2007}).

\bibitem[{\citenamefont{Woods et~al.}(2014)\citenamefont{Woods, Britnell,
  Eckmann, Ma, Lu, Guo, Lin, Yu, Cao, Gorbachev et~al.}}]{wo14}
\bibinfo{author}{\bibfnamefont{C.~R.} \bibnamefont{Woods}},
  \bibinfo{author}{\bibfnamefont{L.}~\bibnamefont{Britnell}},
  \bibinfo{author}{\bibfnamefont{A.}~\bibnamefont{Eckmann}},
  \bibinfo{author}{\bibfnamefont{R.~S.} \bibnamefont{Ma}},
  \bibinfo{author}{\bibfnamefont{J.~C.} \bibnamefont{Lu}},
  \bibinfo{author}{\bibfnamefont{H.~M.} \bibnamefont{Guo}},
  \bibinfo{author}{\bibfnamefont{X.}~\bibnamefont{Lin}},
  \bibinfo{author}{\bibfnamefont{G.~L.} \bibnamefont{Yu}},
  \bibinfo{author}{\bibfnamefont{Y.}~\bibnamefont{Cao}},
  \bibinfo{author}{\bibfnamefont{R.~V.} \bibnamefont{Gorbachev}},
  \bibnamefont{et~al.}, \bibinfo{journal}{Nature Phys.}
  \textbf{\bibinfo{volume}{10}}, \bibinfo{pages}{451} (\bibinfo{year}{2014}).

\bibitem[{\citenamefont{Meunier et~al.}(2016)\citenamefont{Meunier,
  Souza~Filho, Barros, and Dresselhaus}}]{me16}
\bibinfo{author}{\bibfnamefont{V.}~\bibnamefont{Meunier}},
  \bibinfo{author}{\bibfnamefont{A.~G.} \bibnamefont{Souza~Filho}},
  \bibinfo{author}{\bibfnamefont{E.~B.} \bibnamefont{Barros}},
  \bibnamefont{and} \bibinfo{author}{\bibfnamefont{M.~S.}
  \bibnamefont{Dresselhaus}}, \bibinfo{journal}{Rev. Mod. Phys.}
  \textbf{\bibinfo{volume}{88}}, \bibinfo{pages}{025005}
  (\bibinfo{year}{2016}).

\bibitem[{\citenamefont{Castro~Neto et~al.}(2009)\citenamefont{Castro~Neto,
  Guinea, Peres, Novoselov, and Geim}}]{ca09b}
\bibinfo{author}{\bibfnamefont{A.~H.} \bibnamefont{Castro~Neto}},
  \bibinfo{author}{\bibfnamefont{F.}~\bibnamefont{Guinea}},
  \bibinfo{author}{\bibfnamefont{N.~M.~R.} \bibnamefont{Peres}},
  \bibinfo{author}{\bibfnamefont{K.~S.} \bibnamefont{Novoselov}},
  \bibnamefont{and} \bibinfo{author}{\bibfnamefont{A.~K.} \bibnamefont{Geim}},
  \bibinfo{journal}{Rev. Mod. Phys.} \textbf{\bibinfo{volume}{81}},
  \bibinfo{pages}{109} (\bibinfo{year}{2009}).

\bibitem[{\citenamefont{Cao et~al.}(2018)\citenamefont{Cao, Fatemi, Fang,
  Watanabe, Taniguchi, Kaxiras, and Jarillo-Herrero}}]{ca18}
\bibinfo{author}{\bibfnamefont{Y.}~\bibnamefont{Cao}},
  \bibinfo{author}{\bibfnamefont{V.}~\bibnamefont{Fatemi}},
  \bibinfo{author}{\bibfnamefont{S.}~\bibnamefont{Fang}},
  \bibinfo{author}{\bibfnamefont{K.}~\bibnamefont{Watanabe}},
  \bibinfo{author}{\bibfnamefont{T.}~\bibnamefont{Taniguchi}},
  \bibinfo{author}{\bibfnamefont{E.}~\bibnamefont{Kaxiras}}, \bibnamefont{and}
  \bibinfo{author}{\bibfnamefont{P.}~\bibnamefont{Jarillo-Herrero}},
  \bibinfo{journal}{Nature} \textbf{\bibinfo{volume}{556}}, \bibinfo{pages}{43}
  (\bibinfo{year}{2018}).

\bibitem[{\citenamefont{Yankowitz et~al.}(2019)\citenamefont{Yankowitz, Chen,
  Polshyn, Zhang, Watanabe, Taniguchi, Graf, Young, and Dean}}]{ya19}
\bibinfo{author}{\bibfnamefont{M.}~\bibnamefont{Yankowitz}},
  \bibinfo{author}{\bibfnamefont{S.}~\bibnamefont{Chen}},
  \bibinfo{author}{\bibfnamefont{H.}~\bibnamefont{Polshyn}},
  \bibinfo{author}{\bibfnamefont{Y.}~\bibnamefont{Zhang}},
  \bibinfo{author}{\bibfnamefont{K.}~\bibnamefont{Watanabe}},
  \bibinfo{author}{\bibfnamefont{T.}~\bibnamefont{Taniguchi}},
  \bibinfo{author}{\bibfnamefont{D.}~\bibnamefont{Graf}},
  \bibinfo{author}{\bibfnamefont{A.~F.} \bibnamefont{Young}}, \bibnamefont{and}
  \bibinfo{author}{\bibfnamefont{C.~R.} \bibnamefont{Dean}},
  \bibinfo{journal}{Science} \textbf{\bibinfo{volume}{363}},
  \bibinfo{pages}{1059} (\bibinfo{year}{2019}).

\bibitem[{\citenamefont{Sofo et~al.}(2007)\citenamefont{Sofo, Chaudhari, and
  Barber}}]{so07}
\bibinfo{author}{\bibfnamefont{J.~O.} \bibnamefont{Sofo}},
  \bibinfo{author}{\bibfnamefont{A.~S.} \bibnamefont{Chaudhari}},
  \bibnamefont{and} \bibinfo{author}{\bibfnamefont{G.~D.}
  \bibnamefont{Barber}}, \bibinfo{journal}{Phys. Rev. B}
  \textbf{\bibinfo{volume}{75}}, \bibinfo{pages}{153401}
  (\bibinfo{year}{2007}).

\bibitem[{\citenamefont{Wen et~al.}(2011)\citenamefont{Wen, Hand, Labet, Yang,
  Hoffmann, Ashcroft, Oganov, and Lyakhov}}]{we11}
\bibinfo{author}{\bibfnamefont{X.-D.} \bibnamefont{Wen}},
  \bibinfo{author}{\bibfnamefont{L.}~\bibnamefont{Hand}},
  \bibinfo{author}{\bibfnamefont{V.}~\bibnamefont{Labet}},
  \bibinfo{author}{\bibfnamefont{T.}~\bibnamefont{Yang}},
  \bibinfo{author}{\bibfnamefont{R.}~\bibnamefont{Hoffmann}},
  \bibinfo{author}{\bibfnamefont{N.~W.} \bibnamefont{Ashcroft}},
  \bibinfo{author}{\bibfnamefont{A.~R.} \bibnamefont{Oganov}},
  \bibnamefont{and} \bibinfo{author}{\bibfnamefont{A.~O.}
  \bibnamefont{Lyakhov}}, \bibinfo{journal}{PNAS USA}
  \textbf{\bibinfo{volume}{108}}, \bibinfo{pages}{6833} (\bibinfo{year}{2011}).

\bibitem[{\citenamefont{Cadelano et~al.}(2010)\citenamefont{Cadelano, Palla,
  Giordano, and Colombo}}]{ca10}
\bibinfo{author}{\bibfnamefont{E.}~\bibnamefont{Cadelano}},
  \bibinfo{author}{\bibfnamefont{P.~L.} \bibnamefont{Palla}},
  \bibinfo{author}{\bibfnamefont{S.}~\bibnamefont{Giordano}}, \bibnamefont{and}
  \bibinfo{author}{\bibfnamefont{L.}~\bibnamefont{Colombo}},
  \bibinfo{journal}{Phys. Rev. B} \textbf{\bibinfo{volume}{82}},
  \bibinfo{pages}{235414} (\bibinfo{year}{2010}).

\bibitem[{\citenamefont{Elias et~al.}(2009)\citenamefont{Elias, Nair,
  Mohiuddin, Morozov, Blake, Halsall, Ferrari, Boukhvalov, Katsnelson, Geim
  et~al.}}]{el09}
\bibinfo{author}{\bibfnamefont{D.~C.} \bibnamefont{Elias}},
  \bibinfo{author}{\bibfnamefont{R.~R.} \bibnamefont{Nair}},
  \bibinfo{author}{\bibfnamefont{T.~M.~G.} \bibnamefont{Mohiuddin}},
  \bibinfo{author}{\bibfnamefont{S.~V.} \bibnamefont{Morozov}},
  \bibinfo{author}{\bibfnamefont{P.}~\bibnamefont{Blake}},
  \bibinfo{author}{\bibfnamefont{M.~P.} \bibnamefont{Halsall}},
  \bibinfo{author}{\bibfnamefont{A.~C.} \bibnamefont{Ferrari}},
  \bibinfo{author}{\bibfnamefont{D.~W.} \bibnamefont{Boukhvalov}},
  \bibinfo{author}{\bibfnamefont{M.~I.} \bibnamefont{Katsnelson}},
  \bibinfo{author}{\bibfnamefont{A.~K.} \bibnamefont{Geim}},
  \bibnamefont{et~al.}, \bibinfo{journal}{Science}
  \textbf{\bibinfo{volume}{323}}, \bibinfo{pages}{610} (\bibinfo{year}{2009}).

\bibitem[{\citenamefont{Sahin et~al.}(2010)\citenamefont{Sahin, Ataca, and
  Ciraci}}]{sa10b}
\bibinfo{author}{\bibfnamefont{H.}~\bibnamefont{Sahin}},
  \bibinfo{author}{\bibfnamefont{C.}~\bibnamefont{Ataca}}, \bibnamefont{and}
  \bibinfo{author}{\bibfnamefont{S.}~\bibnamefont{Ciraci}},
  \bibinfo{journal}{Phys. Rev. B} \textbf{\bibinfo{volume}{81}},
  \bibinfo{pages}{205417} (\bibinfo{year}{2010}).

\bibitem[{\citenamefont{Wang et~al.}(2016{\natexlab{a}})\citenamefont{Wang,
  Sofer, Bousa, Sedmidubsky, Huber, Matejkova, Michalcova, and Pumera}}]{wa16b}
\bibinfo{author}{\bibfnamefont{L.}~\bibnamefont{Wang}},
  \bibinfo{author}{\bibfnamefont{Z.}~\bibnamefont{Sofer}},
  \bibinfo{author}{\bibfnamefont{D.}~\bibnamefont{Bousa}},
  \bibinfo{author}{\bibfnamefont{D.}~\bibnamefont{Sedmidubsky}},
  \bibinfo{author}{\bibfnamefont{S.}~\bibnamefont{Huber}},
  \bibinfo{author}{\bibfnamefont{S.}~\bibnamefont{Matejkova}},
  \bibinfo{author}{\bibfnamefont{A.}~\bibnamefont{Michalcova}},
  \bibnamefont{and} \bibinfo{author}{\bibfnamefont{M.}~\bibnamefont{Pumera}},
  \bibinfo{journal}{Andgewandte Chemie Intern. Ed.}
  \textbf{\bibinfo{volume}{55}}, \bibinfo{pages}{13965}
  (\bibinfo{year}{2016}{\natexlab{a}}).

\bibitem[{\citenamefont{Mapasha et~al.}(2017)\citenamefont{Mapasha, Molepo, and
  Chetty}}]{ma17b}
\bibinfo{author}{\bibfnamefont{R.~E.} \bibnamefont{Mapasha}},
  \bibinfo{author}{\bibfnamefont{M.~P.} \bibnamefont{Molepo}},
  \bibnamefont{and} \bibinfo{author}{\bibfnamefont{N.}~\bibnamefont{Chetty}},
  \bibinfo{journal}{RSC Adv.} \textbf{\bibinfo{volume}{7}},
  \bibinfo{pages}{39748} (\bibinfo{year}{2017}).

\bibitem[{\citenamefont{Eng et~al.}(2013)\citenamefont{Eng, Poh, Sanek,
  Marysko, Matejkova, Sofer, and Pumera}}]{en13}
\bibinfo{author}{\bibfnamefont{A.~Y.~S.} \bibnamefont{Eng}},
  \bibinfo{author}{\bibfnamefont{H.~L.} \bibnamefont{Poh}},
  \bibinfo{author}{\bibfnamefont{F.}~\bibnamefont{Sanek}},
  \bibinfo{author}{\bibfnamefont{M.}~\bibnamefont{Marysko}},
  \bibinfo{author}{\bibfnamefont{S.}~\bibnamefont{Matejkova}},
  \bibinfo{author}{\bibfnamefont{Z.}~\bibnamefont{Sofer}}, \bibnamefont{and}
  \bibinfo{author}{\bibfnamefont{M.}~\bibnamefont{Pumera}},
  \bibinfo{journal}{ACS Nano} \textbf{\bibinfo{volume}{7}},
  \bibinfo{pages}{5930} (\bibinfo{year}{2013}).

\bibitem[{\citenamefont{Safran}(1994)}]{sa94}
\bibinfo{author}{\bibfnamefont{S.~A.} \bibnamefont{Safran}},
  \emph{\bibinfo{title}{Statistical Thermodynamics of Surfaces, Interfaces, and
  Membranes}} (\bibinfo{publisher}{Addison Wesley}, \bibinfo{address}{New
  York}, \bibinfo{year}{1994}).

\bibitem[{\citenamefont{Nelson et~al.}(2004)\citenamefont{Nelson, Piran, and
  Weinberg}}]{ne04}
\bibinfo{author}{\bibfnamefont{D.}~\bibnamefont{Nelson}},
  \bibinfo{author}{\bibfnamefont{T.}~\bibnamefont{Piran}}, \bibnamefont{and}
  \bibinfo{author}{\bibfnamefont{S.}~\bibnamefont{Weinberg}},
  \emph{\bibinfo{title}{Statistical Mechanics of Membranes and Surfaces}}
  (\bibinfo{publisher}{World Scientific}, \bibinfo{address}{London},
  \bibinfo{year}{2004}).

\bibitem[{\citenamefont{Tarazona et~al.}(2013)\citenamefont{Tarazona, Chac\'on,
  and Bresme}}]{ta13}
\bibinfo{author}{\bibfnamefont{P.}~\bibnamefont{Tarazona}},
  \bibinfo{author}{\bibfnamefont{E.}~\bibnamefont{Chac\'on}}, \bibnamefont{and}
  \bibinfo{author}{\bibfnamefont{F.}~\bibnamefont{Bresme}},
  \bibinfo{journal}{J. Chem. Phys.} \textbf{\bibinfo{volume}{139}},
  \bibinfo{pages}{094902} (\bibinfo{year}{2013}).

\bibitem[{\citenamefont{Fournier and Barbetta}(2008)}]{fo08}
\bibinfo{author}{\bibfnamefont{J.-B.} \bibnamefont{Fournier}} \bibnamefont{and}
  \bibinfo{author}{\bibfnamefont{C.}~\bibnamefont{Barbetta}},
  \bibinfo{journal}{Phys. Rev. Lett.} \textbf{\bibinfo{volume}{100}},
  \bibinfo{pages}{078103} (\bibinfo{year}{2008}).

\bibitem[{\citenamefont{Pop et~al.}(2012)\citenamefont{Pop, Varshney, and
  Roy}}]{po12b}
\bibinfo{author}{\bibfnamefont{E.}~\bibnamefont{Pop}},
  \bibinfo{author}{\bibfnamefont{V.}~\bibnamefont{Varshney}}, \bibnamefont{and}
  \bibinfo{author}{\bibfnamefont{A.~K.} \bibnamefont{Roy}},
  \bibinfo{journal}{MRS Bull.} \textbf{\bibinfo{volume}{37}},
  \bibinfo{pages}{1273} (\bibinfo{year}{2012}).

\bibitem[{\citenamefont{Fong et~al.}(2013)\citenamefont{Fong, Wollman, Ravi,
  Chen, Clerk, Shaw, Leduc, and Schwab}}]{fo13}
\bibinfo{author}{\bibfnamefont{K.~C.} \bibnamefont{Fong}},
  \bibinfo{author}{\bibfnamefont{E.~E.} \bibnamefont{Wollman}},
  \bibinfo{author}{\bibfnamefont{H.}~\bibnamefont{Ravi}},
  \bibinfo{author}{\bibfnamefont{W.}~\bibnamefont{Chen}},
  \bibinfo{author}{\bibfnamefont{A.~A.} \bibnamefont{Clerk}},
  \bibinfo{author}{\bibfnamefont{M.~D.} \bibnamefont{Shaw}},
  \bibinfo{author}{\bibfnamefont{H.~G.} \bibnamefont{Leduc}}, \bibnamefont{and}
  \bibinfo{author}{\bibfnamefont{K.~C.} \bibnamefont{Schwab}},
  \bibinfo{journal}{Phys. Rev. X} \textbf{\bibinfo{volume}{3}},
  \bibinfo{pages}{041008} (\bibinfo{year}{2013}).

\bibitem[{\citenamefont{Wang et~al.}(2016{\natexlab{b}})\citenamefont{Wang,
  Gao, and Huang}}]{wa16}
\bibinfo{author}{\bibfnamefont{P.}~\bibnamefont{Wang}},
  \bibinfo{author}{\bibfnamefont{W.}~\bibnamefont{Gao}}, \bibnamefont{and}
  \bibinfo{author}{\bibfnamefont{R.}~\bibnamefont{Huang}}, \bibinfo{journal}{J.
  Appl. Phys.} \textbf{\bibinfo{volume}{119}}, \bibinfo{pages}{074305}
  (\bibinfo{year}{2016}{\natexlab{b}}).

\bibitem[{\citenamefont{Herrero and Ram\'irez}(2018{\natexlab{a}})}]{he18}
\bibinfo{author}{\bibfnamefont{C.~P.} \bibnamefont{Herrero}} \bibnamefont{and}
  \bibinfo{author}{\bibfnamefont{R.}~\bibnamefont{Ram\'irez}},
  \bibinfo{journal}{J. Chem. Phys.} \textbf{\bibinfo{volume}{148}},
  \bibinfo{pages}{102302} (\bibinfo{year}{2018}{\natexlab{a}}).

\bibitem[{\citenamefont{Costamagna et~al.}(2012)\citenamefont{Costamagna,
  Neek-Amal, Los, and Peeters}}]{co12}
\bibinfo{author}{\bibfnamefont{S.}~\bibnamefont{Costamagna}},
  \bibinfo{author}{\bibfnamefont{M.}~\bibnamefont{Neek-Amal}},
  \bibinfo{author}{\bibfnamefont{J.~H.} \bibnamefont{Los}}, \bibnamefont{and}
  \bibinfo{author}{\bibfnamefont{F.~M.} \bibnamefont{Peeters}},
  \bibinfo{journal}{Phys. Rev. B} \textbf{\bibinfo{volume}{86}},
  \bibinfo{pages}{041408} (\bibinfo{year}{2012}).

\bibitem[{\citenamefont{Gao and Huang}(2014)}]{ga14}
\bibinfo{author}{\bibfnamefont{W.}~\bibnamefont{Gao}} \bibnamefont{and}
  \bibinfo{author}{\bibfnamefont{R.}~\bibnamefont{Huang}}, \bibinfo{journal}{J.
  Mech. Phys. Solids} \textbf{\bibinfo{volume}{66}}, \bibinfo{pages}{42}
  (\bibinfo{year}{2014}).

\bibitem[{\citenamefont{Leenaerts et~al.}(2010)\citenamefont{Leenaerts,
  Peelaers, Hernandez-Nieves, Partoens, and Peeters}}]{le10}
\bibinfo{author}{\bibfnamefont{O.}~\bibnamefont{Leenaerts}},
  \bibinfo{author}{\bibfnamefont{H.}~\bibnamefont{Peelaers}},
  \bibinfo{author}{\bibfnamefont{A.~D.} \bibnamefont{Hernandez-Nieves}},
  \bibinfo{author}{\bibfnamefont{B.}~\bibnamefont{Partoens}}, \bibnamefont{and}
  \bibinfo{author}{\bibfnamefont{F.~M.} \bibnamefont{Peeters}},
  \bibinfo{journal}{Phys. Rev. B} \textbf{\bibinfo{volume}{82}},
  \bibinfo{pages}{195436} (\bibinfo{year}{2010}).

\bibitem[{\citenamefont{Zhou et~al.}(2013)\citenamefont{Zhou, Huang, Chen, and
  Lu}}]{zh13}
\bibinfo{author}{\bibfnamefont{X.~H.} \bibnamefont{Zhou}},
  \bibinfo{author}{\bibfnamefont{Y.}~\bibnamefont{Huang}},
  \bibinfo{author}{\bibfnamefont{X.~S.} \bibnamefont{Chen}}, \bibnamefont{and}
  \bibinfo{author}{\bibfnamefont{W.}~\bibnamefont{Lu}}, \bibinfo{journal}{Solid
  State Commun.} \textbf{\bibinfo{volume}{157}}, \bibinfo{pages}{24}
  (\bibinfo{year}{2013}).

\bibitem[{\citenamefont{Chechin et~al.}(2014)\citenamefont{Chechin, Dmitriev,
  Lobzenko, and Ryabov}}]{ch14}
\bibinfo{author}{\bibfnamefont{G.~M.} \bibnamefont{Chechin}},
  \bibinfo{author}{\bibfnamefont{S.~V.} \bibnamefont{Dmitriev}},
  \bibinfo{author}{\bibfnamefont{I.~P.} \bibnamefont{Lobzenko}},
  \bibnamefont{and} \bibinfo{author}{\bibfnamefont{D.~S.}
  \bibnamefont{Ryabov}}, \bibinfo{journal}{Phys. Rev. B}
  \textbf{\bibinfo{volume}{90}}, \bibinfo{pages}{045432}
  (\bibinfo{year}{2014}).

\bibitem[{\citenamefont{Liu et~al.}({2013})\citenamefont{Liu, Baimova,
  Dmitriev, Wang, Zhu, and Zhou}}]{li13}
\bibinfo{author}{\bibfnamefont{B.}~\bibnamefont{Liu}},
  \bibinfo{author}{\bibfnamefont{J.~A.} \bibnamefont{Baimova}},
  \bibinfo{author}{\bibfnamefont{S.~V.} \bibnamefont{Dmitriev}},
  \bibinfo{author}{\bibfnamefont{X.}~\bibnamefont{Wang}},
  \bibinfo{author}{\bibfnamefont{H.}~\bibnamefont{Zhu}}, \bibnamefont{and}
  \bibinfo{author}{\bibfnamefont{K.}~\bibnamefont{Zhou}}, \bibinfo{journal}{J.
  Phys. D: Appl. Phys.} \textbf{\bibinfo{volume}{{46}}},
  \bibinfo{pages}{{305302}} (\bibinfo{year}{{2013}}).

\bibitem[{\citenamefont{Gillan}(1988)}]{gi88}
\bibinfo{author}{\bibfnamefont{M.~J.} \bibnamefont{Gillan}},
  \bibinfo{journal}{Phil. Mag. A} \textbf{\bibinfo{volume}{58}},
  \bibinfo{pages}{257} (\bibinfo{year}{1988}).

\bibitem[{\citenamefont{Ceperley}(1995)}]{ce95}
\bibinfo{author}{\bibfnamefont{D.~M.} \bibnamefont{Ceperley}},
  \bibinfo{journal}{Rev. Mod. Phys.} \textbf{\bibinfo{volume}{67}},
  \bibinfo{pages}{279} (\bibinfo{year}{1995}).

\bibitem[{\citenamefont{Herrero and Ram\'{\i}rez}(1995)}]{he95}
\bibinfo{author}{\bibfnamefont{C.~P.} \bibnamefont{Herrero}} \bibnamefont{and}
  \bibinfo{author}{\bibfnamefont{R.}~\bibnamefont{Ram\'{\i}rez}},
  \bibinfo{journal}{Phys. Rev. B} \textbf{\bibinfo{volume}{51}},
  \bibinfo{pages}{16761} (\bibinfo{year}{1995}).

\bibitem[{\citenamefont{Ram\'irez and Herrero}(2011)}]{ra11}
\bibinfo{author}{\bibfnamefont{R.}~\bibnamefont{Ram\'irez}} \bibnamefont{and}
  \bibinfo{author}{\bibfnamefont{C.~P.} \bibnamefont{Herrero}},
  \bibinfo{journal}{Phys. Rev. B} \textbf{\bibinfo{volume}{84}},
  \bibinfo{pages}{064130} (\bibinfo{year}{2011}).

\bibitem[{\citenamefont{Los et~al.}(2016)\citenamefont{Los, Fasolino, and
  Katsnelson}}]{lo16}
\bibinfo{author}{\bibfnamefont{J.~H.} \bibnamefont{Los}},
  \bibinfo{author}{\bibfnamefont{A.}~\bibnamefont{Fasolino}}, \bibnamefont{and}
  \bibinfo{author}{\bibfnamefont{M.~I.} \bibnamefont{Katsnelson}},
  \bibinfo{journal}{Phys. Rev. Lett.} \textbf{\bibinfo{volume}{116}},
  \bibinfo{pages}{015901} (\bibinfo{year}{2016}).

\bibitem[{\citenamefont{Herrero and Ram\'irez}(2000)}]{he00c}
\bibinfo{author}{\bibfnamefont{C.~P.} \bibnamefont{Herrero}} \bibnamefont{and}
  \bibinfo{author}{\bibfnamefont{R.}~\bibnamefont{Ram\'irez}},
  \bibinfo{journal}{Phys. Rev. B} \textbf{\bibinfo{volume}{63}},
  \bibinfo{pages}{024103} (\bibinfo{year}{2000}).

\bibitem[{\citenamefont{Ram\'{\i}rez et~al.}(2006)\citenamefont{Ram\'{\i}rez,
  Herrero, and Hern\'andez}}]{ra06}
\bibinfo{author}{\bibfnamefont{R.}~\bibnamefont{Ram\'{\i}rez}},
  \bibinfo{author}{\bibfnamefont{C.~P.} \bibnamefont{Herrero}},
  \bibnamefont{and} \bibinfo{author}{\bibfnamefont{E.~R.}
  \bibnamefont{Hern\'andez}}, \bibinfo{journal}{Phys. Rev. B}
  \textbf{\bibinfo{volume}{73}}, \bibinfo{pages}{245202}
  (\bibinfo{year}{2006}).

\bibitem[{\citenamefont{Herrero and Ram\'{\i}rez}(2007)}]{he07}
\bibinfo{author}{\bibfnamefont{C.~P.} \bibnamefont{Herrero}} \bibnamefont{and}
  \bibinfo{author}{\bibfnamefont{R.}~\bibnamefont{Ram\'{\i}rez}},
  \bibinfo{journal}{Phys. Rev. Lett.} \textbf{\bibinfo{volume}{99}},
  \bibinfo{pages}{205504} (\bibinfo{year}{2007}).

\bibitem[{\citenamefont{Brito et~al.}(2015)\citenamefont{Brito, C\^andido, Hai,
  and Peeters}}]{br15}
\bibinfo{author}{\bibfnamefont{B.~G.~A.} \bibnamefont{Brito}},
  \bibinfo{author}{\bibfnamefont{L.}~\bibnamefont{C\^andido}},
  \bibinfo{author}{\bibfnamefont{G.-Q.} \bibnamefont{Hai}}, \bibnamefont{and}
  \bibinfo{author}{\bibfnamefont{F.~M.} \bibnamefont{Peeters}},
  \bibinfo{journal}{Phys. Rev. B} \textbf{\bibinfo{volume}{92}},
  \bibinfo{pages}{195416} (\bibinfo{year}{2015}).

\bibitem[{\citenamefont{Hasik et~al.}({2018})\citenamefont{Hasik, Tosatti, and
  Martonak}}]{ha18}
\bibinfo{author}{\bibfnamefont{J.}~\bibnamefont{Hasik}},
  \bibinfo{author}{\bibfnamefont{E.}~\bibnamefont{Tosatti}}, \bibnamefont{and}
  \bibinfo{author}{\bibfnamefont{R.}~\bibnamefont{Martonak}},
  \bibinfo{journal}{Phys. Rev. B} \textbf{\bibinfo{volume}{{97}}},
  \bibinfo{pages}{{140301}} (\bibinfo{year}{{2018}}).

\bibitem[{\citenamefont{Herrero and Ram\'irez}(2016)}]{he16}
\bibinfo{author}{\bibfnamefont{C.~P.} \bibnamefont{Herrero}} \bibnamefont{and}
  \bibinfo{author}{\bibfnamefont{R.}~\bibnamefont{Ram\'irez}},
  \bibinfo{journal}{J. Chem. Phys.} \textbf{\bibinfo{volume}{145}},
  \bibinfo{pages}{224701} (\bibinfo{year}{2016}).

\bibitem[{\citenamefont{Davidson et~al.}(2014)\citenamefont{Davidson, Klimes,
  Alfe, and Michaelides}}]{da14}
\bibinfo{author}{\bibfnamefont{E.~R.~M.} \bibnamefont{Davidson}},
  \bibinfo{author}{\bibfnamefont{J.}~\bibnamefont{Klimes}},
  \bibinfo{author}{\bibfnamefont{D.}~\bibnamefont{Alfe}}, \bibnamefont{and}
  \bibinfo{author}{\bibfnamefont{A.}~\bibnamefont{Michaelides}},
  \bibinfo{journal}{ACS Nano} \textbf{\bibinfo{volume}{8}},
  \bibinfo{pages}{9905} (\bibinfo{year}{2014}).

\bibitem[{\citenamefont{Herrero and Ram\'irez}(2009)}]{he09a}
\bibinfo{author}{\bibfnamefont{C.~P.} \bibnamefont{Herrero}} \bibnamefont{and}
  \bibinfo{author}{\bibfnamefont{R.}~\bibnamefont{Ram\'irez}},
  \bibinfo{journal}{Phys. Rev. B} \textbf{\bibinfo{volume}{79}},
  \bibinfo{pages}{115429} (\bibinfo{year}{2009}).

\bibitem[{\citenamefont{Huang and Zeng}({2013})}]{hu13b}
\bibinfo{author}{\bibfnamefont{L.~F.} \bibnamefont{Huang}} \bibnamefont{and}
  \bibinfo{author}{\bibfnamefont{Z.}~\bibnamefont{Zeng}}, \bibinfo{journal}{J.
  Appl. Phys.} \textbf{\bibinfo{volume}{{113}}}, \bibinfo{pages}{{083524}}
  (\bibinfo{year}{{2013}}).

\bibitem[{\citenamefont{Feynman}(1972)}]{fe72}
\bibinfo{author}{\bibfnamefont{R.~P.} \bibnamefont{Feynman}},
  \emph{\bibinfo{title}{Statistical Mechanics}}
  (\bibinfo{publisher}{Addison-Wesley}, \bibinfo{address}{New York},
  \bibinfo{year}{1972}).

\bibitem[{\citenamefont{Kleinert}(1990)}]{kl90}
\bibinfo{author}{\bibfnamefont{H.}~\bibnamefont{Kleinert}},
  \emph{\bibinfo{title}{Path Integrals in Quantum Mechanics, Statistics and
  Polymer Physics}} (\bibinfo{publisher}{World Scientific},
  \bibinfo{address}{Singapore}, \bibinfo{year}{1990}).

\bibitem[{\citenamefont{Chandler and Wolynes}(1981)}]{ch81}
\bibinfo{author}{\bibfnamefont{D.}~\bibnamefont{Chandler}} \bibnamefont{and}
  \bibinfo{author}{\bibfnamefont{P.~G.} \bibnamefont{Wolynes}},
  \bibinfo{journal}{J. Chem. Phys.} \textbf{\bibinfo{volume}{74}},
  \bibinfo{pages}{4078} (\bibinfo{year}{1981}).

\bibitem[{\citenamefont{Tuckerman}(2010)}]{tu10}
\bibinfo{author}{\bibfnamefont{M.~E.} \bibnamefont{Tuckerman}},
  \emph{\bibinfo{title}{Statistical Mechanics: Theory and Molecular
  Simulation}} (\bibinfo{publisher}{Oxford University Press},
  \bibinfo{address}{Oxford}, \bibinfo{year}{2010}).

\bibitem[{\citenamefont{Herrero and Ram\'irez}(2014)}]{he14}
\bibinfo{author}{\bibfnamefont{C.~P.} \bibnamefont{Herrero}} \bibnamefont{and}
  \bibinfo{author}{\bibfnamefont{R.}~\bibnamefont{Ram\'irez}},
  \bibinfo{journal}{J. Phys.: Condens. Matter} \textbf{\bibinfo{volume}{26}},
  \bibinfo{pages}{233201} (\bibinfo{year}{2014}).

\bibitem[{\citenamefont{Ram\'irez and Herrero}(2017)}]{ra17}
\bibinfo{author}{\bibfnamefont{R.}~\bibnamefont{Ram\'irez}} \bibnamefont{and}
  \bibinfo{author}{\bibfnamefont{C.~P.} \bibnamefont{Herrero}},
  \bibinfo{journal}{Phys. Rev. B} \textbf{\bibinfo{volume}{95}},
  \bibinfo{pages}{045423} (\bibinfo{year}{2017}).

\bibitem[{\citenamefont{Shiba et~al.}(2016)\citenamefont{Shiba, Noguchi, and
  Fournier}}]{sh16}
\bibinfo{author}{\bibfnamefont{H.}~\bibnamefont{Shiba}},
  \bibinfo{author}{\bibfnamefont{H.}~\bibnamefont{Noguchi}}, \bibnamefont{and}
  \bibinfo{author}{\bibfnamefont{J.-B.} \bibnamefont{Fournier}},
  \bibinfo{journal}{Soft Matter} \textbf{\bibinfo{volume}{12}},
  \bibinfo{pages}{2373} (\bibinfo{year}{2016}).

\bibitem[{\citenamefont{Herrero and Ram\'irez}(2018{\natexlab{b}})}]{he18b}
\bibinfo{author}{\bibfnamefont{C.~P.} \bibnamefont{Herrero}} \bibnamefont{and}
  \bibinfo{author}{\bibfnamefont{R.}~\bibnamefont{Ram\'irez}},
  \bibinfo{journal}{Phys. Rev. B} \textbf{\bibinfo{volume}{97}},
  \bibinfo{pages}{195433} (\bibinfo{year}{2018}{\natexlab{b}}).

\bibitem[{\citenamefont{Martyna et~al.}(1999)\citenamefont{Martyna, Hughes, and
  Tuckerman}}]{ma99}
\bibinfo{author}{\bibfnamefont{G.~J.} \bibnamefont{Martyna}},
  \bibinfo{author}{\bibfnamefont{A.}~\bibnamefont{Hughes}}, \bibnamefont{and}
  \bibinfo{author}{\bibfnamefont{M.~E.} \bibnamefont{Tuckerman}},
  \bibinfo{journal}{J. Chem. Phys.} \textbf{\bibinfo{volume}{110}},
  \bibinfo{pages}{3275} (\bibinfo{year}{1999}).

\bibitem[{\citenamefont{Martyna et~al.}(1996)\citenamefont{Martyna, Tuckerman,
  Tobias, and Klein}}]{ma96}
\bibinfo{author}{\bibfnamefont{G.~J.} \bibnamefont{Martyna}},
  \bibinfo{author}{\bibfnamefont{M.~E.} \bibnamefont{Tuckerman}},
  \bibinfo{author}{\bibfnamefont{D.~J.} \bibnamefont{Tobias}},
  \bibnamefont{and} \bibinfo{author}{\bibfnamefont{M.~L.} \bibnamefont{Klein}},
  \bibinfo{journal}{Mol. Phys.} \textbf{\bibinfo{volume}{87}},
  \bibinfo{pages}{1117} (\bibinfo{year}{1996}).

\bibitem[{\citenamefont{Herrero et~al.}(2006)\citenamefont{Herrero,
  Ram\'{\i}rez, and Hern\'andez}}]{he06}
\bibinfo{author}{\bibfnamefont{C.~P.} \bibnamefont{Herrero}},
  \bibinfo{author}{\bibfnamefont{R.}~\bibnamefont{Ram\'{\i}rez}},
  \bibnamefont{and} \bibinfo{author}{\bibfnamefont{E.~R.}
  \bibnamefont{Hern\'andez}}, \bibinfo{journal}{Phys. Rev. B}
  \textbf{\bibinfo{volume}{73}}, \bibinfo{pages}{245211}
  (\bibinfo{year}{2006}).

\bibitem[{\citenamefont{Herman et~al.}(1982)\citenamefont{Herman, Bruskin, and
  Berne}}]{he82}
\bibinfo{author}{\bibfnamefont{M.~F.} \bibnamefont{Herman}},
  \bibinfo{author}{\bibfnamefont{E.~J.} \bibnamefont{Bruskin}},
  \bibnamefont{and} \bibinfo{author}{\bibfnamefont{B.~J.} \bibnamefont{Berne}},
  \bibinfo{journal}{J. Chem. Phys.} \textbf{\bibinfo{volume}{76}},
  \bibinfo{pages}{5150} (\bibinfo{year}{1982}).

\bibitem[{\citenamefont{Porezag et~al.}(1995)\citenamefont{Porezag, Frauenheim,
  K\"ohler, Seifert, and Kaschner}}]{po95}
\bibinfo{author}{\bibfnamefont{D.}~\bibnamefont{Porezag}},
  \bibinfo{author}{\bibfnamefont{T.}~\bibnamefont{Frauenheim}},
  \bibinfo{author}{\bibfnamefont{T.}~\bibnamefont{K\"ohler}},
  \bibinfo{author}{\bibfnamefont{G.}~\bibnamefont{Seifert}}, \bibnamefont{and}
  \bibinfo{author}{\bibfnamefont{R.}~\bibnamefont{Kaschner}},
  \bibinfo{journal}{Phys. Rev. B} \textbf{\bibinfo{volume}{51}},
  \bibinfo{pages}{12947} (\bibinfo{year}{1995}).

\bibitem[{\citenamefont{Goringe et~al.}(1997)\citenamefont{Goringe, Bowler, and
  Hern\'andez}}]{go97}
\bibinfo{author}{\bibfnamefont{C.~M.} \bibnamefont{Goringe}},
  \bibinfo{author}{\bibfnamefont{D.~R.} \bibnamefont{Bowler}},
  \bibnamefont{and}
  \bibinfo{author}{\bibfnamefont{E.}~\bibnamefont{Hern\'andez}},
  \bibinfo{journal}{Rep. Prog. Phys.} \textbf{\bibinfo{volume}{60}},
  \bibinfo{pages}{1447} (\bibinfo{year}{1997}).

\bibitem[{\citenamefont{Johnson et~al.}(1993)\citenamefont{Johnson, Gill, and
  Pople}}]{jo93}
\bibinfo{author}{\bibfnamefont{B.~G.} \bibnamefont{Johnson}},
  \bibinfo{author}{\bibfnamefont{P.~M.~W.} \bibnamefont{Gill}},
  \bibnamefont{and} \bibinfo{author}{\bibfnamefont{J.~A.} \bibnamefont{Pople}},
  \bibinfo{journal}{J. Chem. Phys.} \textbf{\bibinfo{volume}{98}},
  \bibinfo{pages}{5612} (\bibinfo{year}{1993}).

\bibitem[{\citenamefont{L{\'o}pez-Ciudad
  et~al.}(2003)\citenamefont{L{\'o}pez-Ciudad, Ram{\'{\i}}rez, Schulte, and
  B\"ohm}}]{lo03}
\bibinfo{author}{\bibfnamefont{T.}~\bibnamefont{L{\'o}pez-Ciudad}},
  \bibinfo{author}{\bibfnamefont{R.}~\bibnamefont{Ram{\'{\i}}rez}},
  \bibinfo{author}{\bibfnamefont{J.}~\bibnamefont{Schulte}}, \bibnamefont{and}
  \bibinfo{author}{\bibfnamefont{M.~C.} \bibnamefont{B\"ohm}},
  \bibinfo{journal}{J. Chem. Phys.} \textbf{\bibinfo{volume}{119}},
  \bibinfo{pages}{4328} (\bibinfo{year}{2003}).

\bibitem[{\citenamefont{B\"ohm et~al.}(2001)\citenamefont{B\"ohm, Schulte,
  Hern\'andez, and Ram{\'{\i}}rez}}]{bo01}
\bibinfo{author}{\bibfnamefont{M.~C.} \bibnamefont{B\"ohm}},
  \bibinfo{author}{\bibfnamefont{J.}~\bibnamefont{Schulte}},
  \bibinfo{author}{\bibfnamefont{E.}~\bibnamefont{Hern\'andez}},
  \bibnamefont{and}
  \bibinfo{author}{\bibfnamefont{R.}~\bibnamefont{Ram{\'{\i}}rez}},
  \bibinfo{journal}{Chem. Phys.} \textbf{\bibinfo{volume}{264}},
  \bibinfo{pages}{371} (\bibinfo{year}{2001}).

\bibitem[{\citenamefont{Herrero and Ram\'irez}(2010)}]{he10b}
\bibinfo{author}{\bibfnamefont{C.~P.} \bibnamefont{Herrero}} \bibnamefont{and}
  \bibinfo{author}{\bibfnamefont{R.}~\bibnamefont{Ram\'irez}},
  \bibinfo{journal}{J. Phys. D: Appl. Phys.} \textbf{\bibinfo{volume}{43}},
  \bibinfo{pages}{255402} (\bibinfo{year}{2010}).

\bibitem[{\citenamefont{Gillan}(1990)}]{gi90}
\bibinfo{author}{\bibfnamefont{M.~J.} \bibnamefont{Gillan}}, in
  \emph{\bibinfo{booktitle}{Computer Modelling of Fluids, Polymers and
  Solids}}, edited by \bibinfo{editor}{\bibfnamefont{C.~R.~A.}
  \bibnamefont{Catlow}}, \bibinfo{editor}{\bibfnamefont{S.~C.}
  \bibnamefont{Parker}}, \bibnamefont{and}
  \bibinfo{editor}{\bibfnamefont{M.~P.} \bibnamefont{Allen}}
  (\bibinfo{publisher}{Kluwer}, \bibinfo{address}{Dordrecht},
  \bibinfo{year}{1990}), p. \bibinfo{pages}{155}.

\bibitem[{\citenamefont{Landau and Lifshitz}(1980)}]{la80}
\bibinfo{author}{\bibfnamefont{L.~D.} \bibnamefont{Landau}} \bibnamefont{and}
  \bibinfo{author}{\bibfnamefont{E.~M.} \bibnamefont{Lifshitz}},
  \emph{\bibinfo{title}{Statistical Physics}} (\bibinfo{publisher}{Pergamon},
  \bibinfo{address}{Oxford}, \bibinfo{year}{1980}), \bibinfo{edition}{3rd} ed.

\bibitem[{\citenamefont{Ram\'irez and Herrero}(2019)}]{ra19}
\bibinfo{author}{\bibfnamefont{R.}~\bibnamefont{Ram\'irez}} \bibnamefont{and}
  \bibinfo{author}{\bibfnamefont{C.~P.} \bibnamefont{Herrero}},
  \bibinfo{journal}{J. Chem. Phys.} \textbf{\bibinfo{volume}{151}},
  \bibinfo{pages}{224107} (\bibinfo{year}{2019}).

\bibitem[{\citenamefont{Landau and Lifshitz}(1965)}]{la65}
\bibinfo{author}{\bibfnamefont{L.~D.} \bibnamefont{Landau}} \bibnamefont{and}
  \bibinfo{author}{\bibfnamefont{E.~M.} \bibnamefont{Lifshitz}},
  \emph{\bibinfo{title}{Quantum Mechanics}} (\bibinfo{publisher}{Pergamon},
  \bibinfo{address}{Oxford}, \bibinfo{year}{1965}), \bibinfo{edition}{2nd} ed.

\bibitem[{\citenamefont{Kittel}(1966)}]{ki66}
\bibinfo{author}{\bibfnamefont{C.}~\bibnamefont{Kittel}},
  \emph{\bibinfo{title}{Introduction to Solid State Physics}}
  (\bibinfo{publisher}{Wiley}, \bibinfo{address}{New York},
  \bibinfo{year}{1966}).

\bibitem[{\citenamefont{Yamanaka et~al.}(1994)\citenamefont{Yamanaka, Morimoto,
  and Kanda}}]{ya94}
\bibinfo{author}{\bibfnamefont{T.}~\bibnamefont{Yamanaka}},
  \bibinfo{author}{\bibfnamefont{S.}~\bibnamefont{Morimoto}}, \bibnamefont{and}
  \bibinfo{author}{\bibfnamefont{H.}~\bibnamefont{Kanda}},
  \bibinfo{journal}{Phys. Rev. B} \textbf{\bibinfo{volume}{49}},
  \bibinfo{pages}{9341} (\bibinfo{year}{1994}).

\bibitem[{\citenamefont{Ramdas et~al.}(1993)\citenamefont{Ramdas, Rodriguez,
  Grimsditch, Anthony, and Banholzer}}]{ra93b}
\bibinfo{author}{\bibfnamefont{A.~K.} \bibnamefont{Ramdas}},
  \bibinfo{author}{\bibfnamefont{S.}~\bibnamefont{Rodriguez}},
  \bibinfo{author}{\bibfnamefont{M.}~\bibnamefont{Grimsditch}},
  \bibinfo{author}{\bibfnamefont{T.~R.} \bibnamefont{Anthony}},
  \bibnamefont{and} \bibinfo{author}{\bibfnamefont{W.~F.}
  \bibnamefont{Banholzer}}, \bibinfo{journal}{Phys. Rev. Lett.}
  \textbf{\bibinfo{volume}{71}}, \bibinfo{pages}{189} (\bibinfo{year}{1993}).

\bibitem[{\citenamefont{Kazimorov et~al.}(1998)\citenamefont{Kazimorov,
  Zegenhagen, and Cardona}}]{ka98}
\bibinfo{author}{\bibfnamefont{A.}~\bibnamefont{Kazimorov}},
  \bibinfo{author}{\bibfnamefont{J.}~\bibnamefont{Zegenhagen}},
  \bibnamefont{and} \bibinfo{author}{\bibfnamefont{M.}~\bibnamefont{Cardona}},
  \bibinfo{journal}{Science} \textbf{\bibinfo{volume}{282}},
  \bibinfo{pages}{930} (\bibinfo{year}{1998}).

\bibitem[{\citenamefont{Ram\'irez et~al.}(2016)\citenamefont{Ram\'irez,
  Chac\'on, and Herrero}}]{ra16}
\bibinfo{author}{\bibfnamefont{R.}~\bibnamefont{Ram\'irez}},
  \bibinfo{author}{\bibfnamefont{E.}~\bibnamefont{Chac\'on}}, \bibnamefont{and}
  \bibinfo{author}{\bibfnamefont{C.~P.} \bibnamefont{Herrero}},
  \bibinfo{journal}{Phys. Rev. B} \textbf{\bibinfo{volume}{93}},
  \bibinfo{pages}{235419} (\bibinfo{year}{2016}).

\bibitem[{\citenamefont{Zakharchenko et~al.}(2009)\citenamefont{Zakharchenko,
  Katsnelson, and Fasolino}}]{za09}
\bibinfo{author}{\bibfnamefont{K.~V.} \bibnamefont{Zakharchenko}},
  \bibinfo{author}{\bibfnamefont{M.~I.} \bibnamefont{Katsnelson}},
  \bibnamefont{and} \bibinfo{author}{\bibfnamefont{A.}~\bibnamefont{Fasolino}},
  \bibinfo{journal}{Phys. Rev. Lett.} \textbf{\bibinfo{volume}{102}},
  \bibinfo{pages}{046808} (\bibinfo{year}{2009}).

\bibitem[{\citenamefont{Callen}(1960)}]{ca60}
\bibinfo{author}{\bibfnamefont{H.~B.} \bibnamefont{Callen}},
  \emph{\bibinfo{title}{Thermodynamics}} (\bibinfo{publisher}{John Wiley},
  \bibinfo{address}{New York}, \bibinfo{year}{1960}).

\bibitem[{\citenamefont{Ruiz-Vargas et~al.}(2011)\citenamefont{Ruiz-Vargas,
  Zhuang, Huang, van~der Zande, Garg, McEuen, Muller, Hennig, and Park}}]{ru11}
\bibinfo{author}{\bibfnamefont{C.~S.} \bibnamefont{Ruiz-Vargas}},
  \bibinfo{author}{\bibfnamefont{H.~L.} \bibnamefont{Zhuang}},
  \bibinfo{author}{\bibfnamefont{P.~Y.} \bibnamefont{Huang}},
  \bibinfo{author}{\bibfnamefont{A.~M.} \bibnamefont{van~der Zande}},
  \bibinfo{author}{\bibfnamefont{S.}~\bibnamefont{Garg}},
  \bibinfo{author}{\bibfnamefont{P.~L.} \bibnamefont{McEuen}},
  \bibinfo{author}{\bibfnamefont{D.~A.} \bibnamefont{Muller}},
  \bibinfo{author}{\bibfnamefont{R.~G.} \bibnamefont{Hennig}},
  \bibnamefont{and} \bibinfo{author}{\bibfnamefont{J.}~\bibnamefont{Park}},
  \bibinfo{journal}{Nano Lett.} \textbf{\bibinfo{volume}{11}},
  \bibinfo{pages}{2259} (\bibinfo{year}{2011}).

\bibitem[{\citenamefont{Kosmrlj and Nelson}(2013)}]{ko13}
\bibinfo{author}{\bibfnamefont{A.}~\bibnamefont{Kosmrlj}} \bibnamefont{and}
  \bibinfo{author}{\bibfnamefont{D.~R.} \bibnamefont{Nelson}},
  \bibinfo{journal}{Phys. Rev. E} \textbf{\bibinfo{volume}{88}},
  \bibinfo{pages}{012136} (\bibinfo{year}{2013}).

\bibitem[{\citenamefont{Kosmrlj and Nelson}(2014)}]{ko14}
\bibinfo{author}{\bibfnamefont{A.}~\bibnamefont{Kosmrlj}} \bibnamefont{and}
  \bibinfo{author}{\bibfnamefont{D.~R.} \bibnamefont{Nelson}},
  \bibinfo{journal}{Phys. Rev. E} \textbf{\bibinfo{volume}{89}},
  \bibinfo{pages}{022126} (\bibinfo{year}{2014}).

\bibitem[{\citenamefont{Seol et~al.}(2010)\citenamefont{Seol, Jo, Moore,
  Lindsay, Aitken, Pettes, Li, Yao, Huang, Broido et~al.}}]{se10}
\bibinfo{author}{\bibfnamefont{J.~H.} \bibnamefont{Seol}},
  \bibinfo{author}{\bibfnamefont{I.}~\bibnamefont{Jo}},
  \bibinfo{author}{\bibfnamefont{A.~L.} \bibnamefont{Moore}},
  \bibinfo{author}{\bibfnamefont{L.}~\bibnamefont{Lindsay}},
  \bibinfo{author}{\bibfnamefont{Z.~H.} \bibnamefont{Aitken}},
  \bibinfo{author}{\bibfnamefont{M.~T.} \bibnamefont{Pettes}},
  \bibinfo{author}{\bibfnamefont{X.}~\bibnamefont{Li}},
  \bibinfo{author}{\bibfnamefont{Z.}~\bibnamefont{Yao}},
  \bibinfo{author}{\bibfnamefont{R.}~\bibnamefont{Huang}},
  \bibinfo{author}{\bibfnamefont{D.}~\bibnamefont{Broido}},
  \bibnamefont{et~al.}, \bibinfo{journal}{Science}
  \textbf{\bibinfo{volume}{328}}, \bibinfo{pages}{213} (\bibinfo{year}{2010}).

\bibitem[{\citenamefont{Prasher}(2010)}]{pr10}
\bibinfo{author}{\bibfnamefont{R.}~\bibnamefont{Prasher}},
  \bibinfo{journal}{Science} \textbf{\bibinfo{volume}{328}},
  \bibinfo{pages}{185} (\bibinfo{year}{2010}).

\bibitem[{\citenamefont{Behroozi}(1996)}]{be96b}
\bibinfo{author}{\bibfnamefont{F.}~\bibnamefont{Behroozi}},
  \bibinfo{journal}{Langmuir} \textbf{\bibinfo{volume}{12}},
  \bibinfo{pages}{2289} (\bibinfo{year}{1996}).

\bibitem[{\citenamefont{Cohen-Tannoudji
  et~al.}(1977)\citenamefont{Cohen-Tannoudji, Liu, and Lal\"oe}}]{co77}
\bibinfo{author}{\bibfnamefont{C.}~\bibnamefont{Cohen-Tannoudji}},
  \bibinfo{author}{\bibfnamefont{B.}~\bibnamefont{Liu}}, \bibnamefont{and}
  \bibinfo{author}{\bibfnamefont{F.}~\bibnamefont{Lal\"oe}},
  \emph{\bibinfo{title}{Quantum Mechanics}}, vol.~\bibinfo{volume}{1}
  (\bibinfo{publisher}{Wiley}, \bibinfo{address}{New York},
  \bibinfo{year}{1977}).

\end{thebibliography}
\end{document}